\documentclass[aps,pra,showpacs,amsmath,amssymb,reprint,10pt,longbibliography,noeprint]{revtex4-1}
\pdfoutput=1

\usepackage{amsmath,amsfonts,amssymb,amsthm,bbm,graphicx,enumerate,times}
\usepackage{mathtools}
\usepackage[usenames,dvipsnames]{color}
\usepackage{todonotes} 
\usepackage{comment}
\usepackage{float}
\usepackage{hyperref}
\usepackage{tikz}
\usepackage{graphicx}
\usepackage{bm}
\usetikzlibrary{shapes,positioning}
\usepackage{etoolbox}

 \newtheorem{theorem}{Theorem}

 \newtheorem{lemma}[theorem]{Lemma}
 
 \newtheorem{definition}[theorem]{Definition}

\definecolor{henrik}{rgb}{.8,.3,0}

\newcommand{\mb}[1]{\mathbb{#1}}

\newcommand{\Tr}{\mathrm{Tr}} 

\newcommand{\id}{\mathbbm{1}}

\renewcommand{\mod}{\mathrm{\, mod\, }}

\newcommand{\1}{\mathrm{id}}

\newcommand{\Z}{\mb{Z}}

\newcommand{\CC}{\mb{C}}

\renewcommand{\1}{\id}

\newcommand{\ket}[1]{\left.\left|{#1}\right.\right\rangle}

\newcommand{\bra}[1]{\left.\left\langle{#1}\right.\right|}

\newcommand{\braket}[2]{\left\langle #1 \middle| #2 \right\rangle}

\newcommand{\ketbra}[2]{\ket{#1} \!\! \bra{#2}}

  \newcommand{\proj}[1]{\ketbra{#1}{#1}}

\renewcommand{\vec}[1]{\pmb{#1}}

\newcommand{\eps}{\varepsilon}

\begin{document}
\title{Quantized and maximum entanglement from sublattice symmetry}
 
\author{Henrik Wilming}
\author{Tobias J. Osborne}
\affiliation{Leibniz Universit\"at Hannover, Appelstra\ss e 2, 30167 Hannover, Germany}
\begin{abstract} 
	We observe that the many-body eigenstates of any quadratic, fermionic Hamiltonian with sublattice symmetry have quantized entanglement entropies between the sublattices: the entanglement comes in multiple singlets. Moreover, such systems always have a ground state that is maximally entangled between the two sublattices. In fact we also show that under the same assumptions there always exists a (potentially distinct) basis of energy eigenstates that do not conserve the particle number in which each energy eigenstate is maximally entangled between the sublattices.
No additional properties, such as translation invariance, are required. 
We also show that the quantization of ground state entanglement may persist when interactions are introduced.
\end{abstract}
\maketitle

\section{Introduction}
{Current progress in quantum many-body physics is largely driven by better understanding of the entanglement structure of their eigenstates.
Prominent examples are the Area Law behavior of ground states~\cite{Eisert2010}, leading to efficient tensor-network representations~\cite{Schollwoeck2011,Cirac2021}, 
topological quantum order  and its connection to quantum error correcting codes \cite{Wen1990,Kitaev2003,Kitaev2006,Wen2013}, and the thermalizing non-equilibrium dynamics of complex quantum systems due to strong entanglement as dictated by the eigenstate thermalization hypothesis \cite{Deutsch1991,Srednicki1994,DAlessio2016,Gogolin2016} or lack thereof due to weakly entangled eigenstates in systems showing many-body localization \cite{Gornyi2005,Basko2006,Nandkishore2015,Abanin2019} or quantum many-body scars \cite{Bernien2017,Turner2018,Turner2018a,Ho2019,Choi2019,Schecter2019,Alhambra2020,Serbyn2021}. 

So far, most research has focussed on understanding the entanglement of a connected subsystem to the remainder of the system.
This leaves open what can be learned from also considering how many disconnected parts of a many-body system are entangled.
For example when we consider a sublattice of a bipartite lattice, then ground states of local Hamiltonians may show extensive entanglement entropies, as was observed for specific models in Refs. \cite{Keating2006,Chen2006,Chen2006a,Igloi2008,Igloi2010,He2017}.
In fact, extensive entanglement entropy for disconnected subsystems is typically expected, even for states that fulfill an area law, such as matrix product states \cite{Rolandi2020,Haferkamp2021}. 
Such extensive entropy in energy eigenstates of disconnected subsystems can have strong consequences. 
For example, suppose if it holds for all energy eigenstates of a generic, interacting many-body system (as expected from the eigenstate thermalization hypothesis). 
Then the system can be proven to equilibrate to high precision from low-entangled initial states \cite{Wilming2019}.
Such results call for systematic studies of the entanglement behaviour of disconnected subsystems in many-body systems. }

\begin{figure}[tb]
	\includegraphics[width=.7\linewidth]{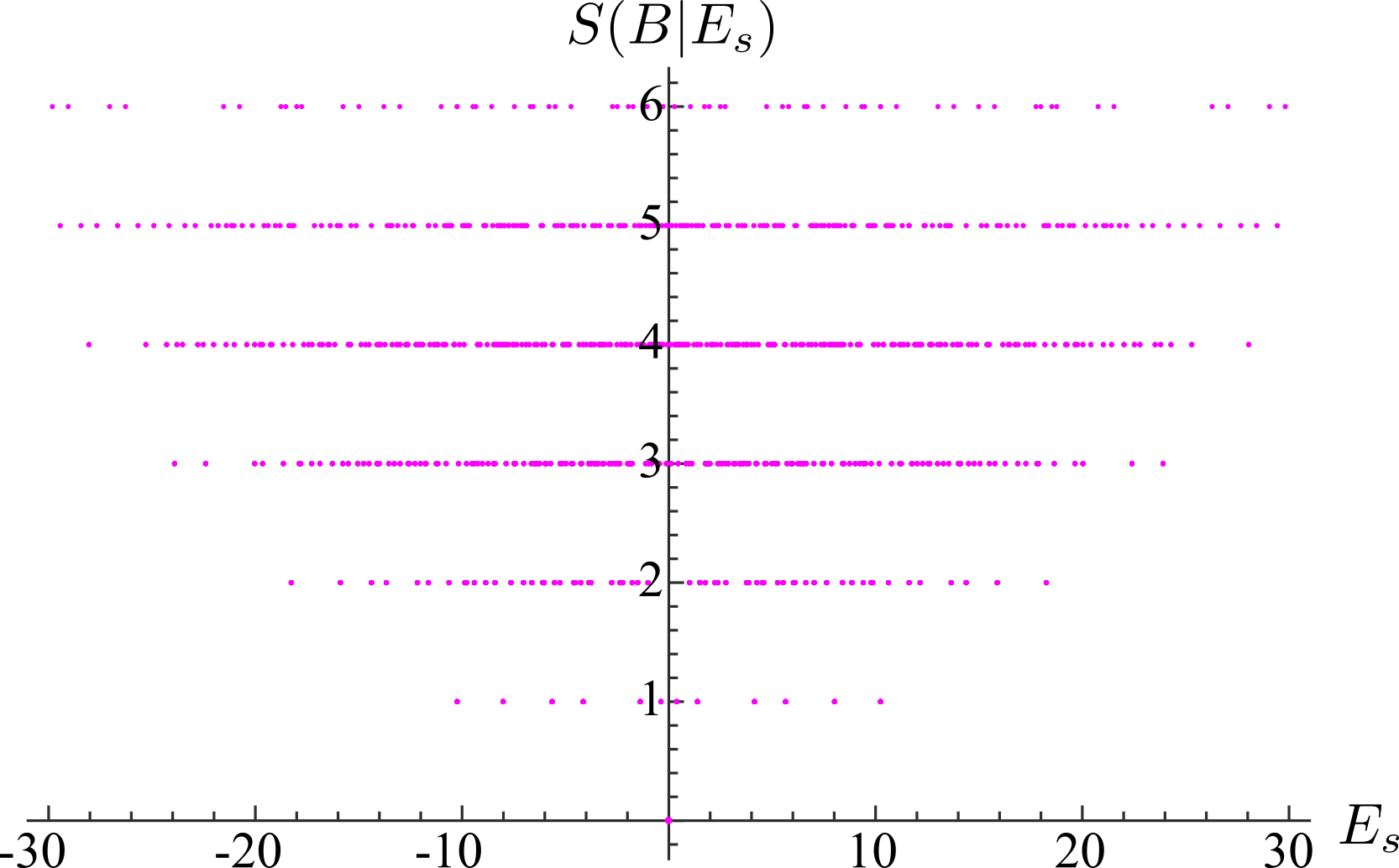}
	\caption{Entanglement entropy of odd sites $S(B|E_s)$ (in units of $\log(2)$) in energy eigenstate $\ket{E_s}$ vs. energy $E_s$ for all $2^{12}$ eigenstates of a randomly chosen Hamiltonian $H_F$ on a chain of $N=12$ sites with $A$ and $B$ corresponding to the even and odd lattice sites, respectively. The quantization of the entropy is clearly visible. }
	\label{fig:spectrum}
\end{figure}

Here, we show that a particularly drastic form of extensive entanglement in ground states appears in a large class of models of non-interacting (quasi-free) fermions, where certain disconnected subsystems are \emph{maximally entangled} with the remainder of the system (we refer to Refs.~\cite{Botero2004,Shapourian2017,Eisert2018} and references therein for discussions on the general entanglement stucture in quasi-free fermionic systems).
The only requirement is that the Hamiltonian features a general form of sublattice symmetry.
This suggests the possibility to generate large-scale, maximally entangled quantum states by cooling suitable fermionic systems to their ground state.
Moreover, we show that the entanglement in all energy eigenstates comes in discrete units of singlets (see Fig.~\ref{fig:spectrum}).
While these observations have been made for the specific case of nearest-neighbor fermionic hopping Hamiltonians on a bipartite lattice before, see e.g. Ref.~\cite{Igloi2010}, we here show that it applies to every quadratic fermionic Hamiltonian with sublattice symmetry.
{Moreover, we present first result indicating that these effects may survive in presense of interactions.}
 
\begin{figure}[tb]
	\includegraphics[width=.7\linewidth]{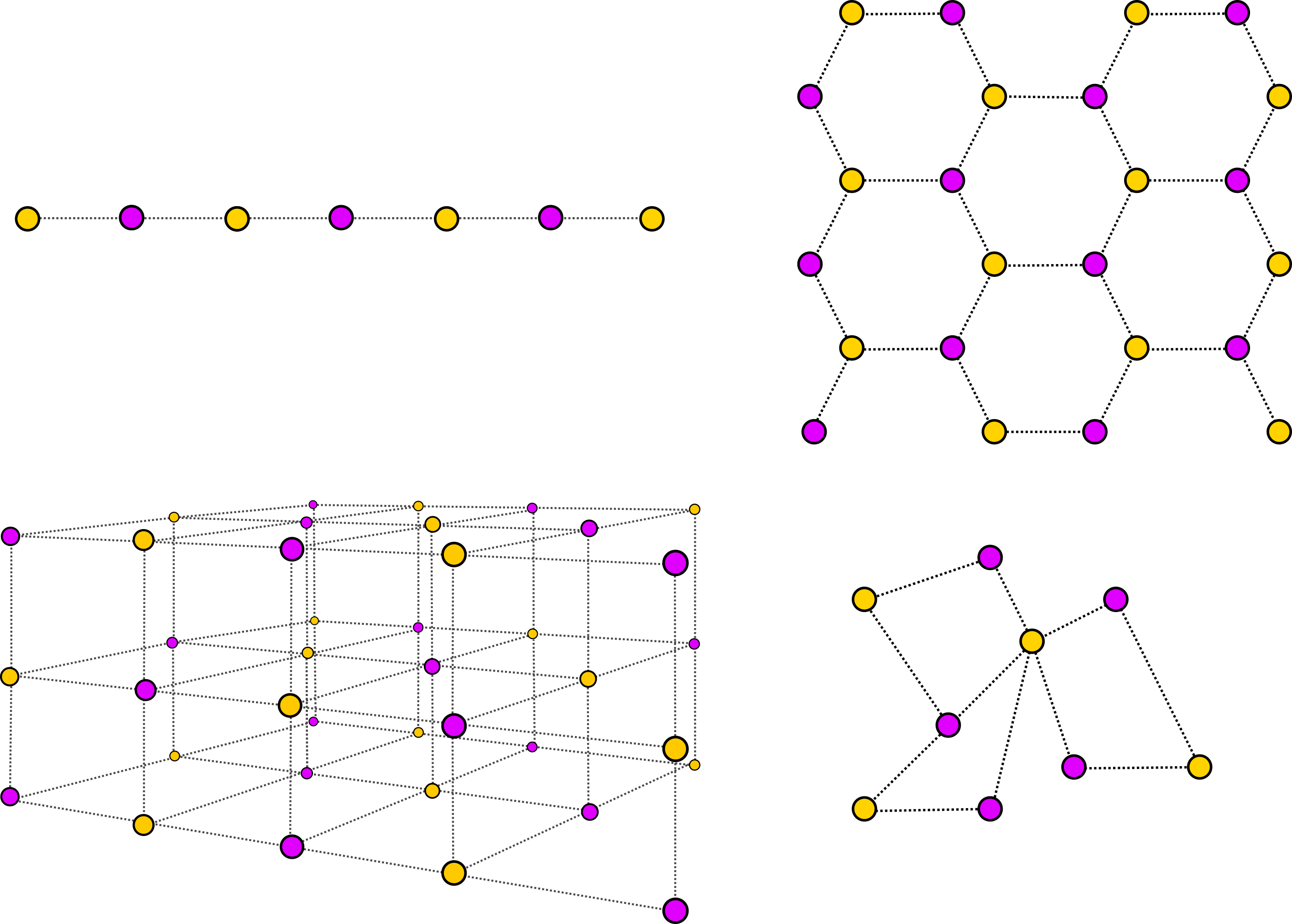}
	\caption{{Exemplary geometries to which our results apply. The only condition is that the fermionic modes can be partitioned into two sets $A$ and $B$, here denoted by yellow/purple dots, and such that the only non-zero tunneling amplitudes (dashed, gray lines) are between the two sets.}}
	\label{fig:geometries}
\end{figure}

\section{Set-up} We consider $N$ fermionic modes described by the annihilation and creation operators $f_i,f_j^\dagger$ {(with the canonical anti-commutation relations $\{f_i,f_j^\dagger\}=\delta_{ij}, \{f_i,f_j\}=0$)} and assume that we can partition the $N$ modes into two disjoint sets $A$ and $B$,
such that the Hamiltonian can be written as
\begin{align}\label{eq:hamiltonian}
	H_F = \sum_{i\in A,j\in B} f_i^\dagger h_{ij} f_j + \mathrm{H.c.},
\end{align}
where $h$ is the single-particle Hamiltonian on the $N$ fermionic modes (below we also discuss the case where the Hamiltonian is not number-conserving).
Such Hamiltonians are ubiquitous as effective descriptions of many-particle systems. 
We will say that $H_F$ has a \emph{sublattice symmetry}, because typically we will have in mind that the $N$ modes correspond to a bipartite lattice
with $A$ and $B$ being the two sublattices. For example, we could consider a one-dimensional chain with $A$ and $B$ being even and odd lattice sites, respectively, and the Hamiltonian given by
\begin{align}\label{eq:XX}
	H_F^{XX} = \frac{1}{2}\sum_{j=1}^{N}  f_j^\dagger f_{j+1} + H.c.
\end{align}
with {open} boundary conditions.
Another example would be a tight-binding model for graphene with only nearest-neighbor tunneling, see Figure~\ref{fig:geometries} for illustrations of possible geometries for which our results apply.
Our definition also allows for multiple fermionic modes per lattice site (for example due to fermions with spin), as long as every mode in one of the sublattices is only coupled to modes in the other sublattice. For example, if we have fermions with spins on a one-dimensional chain and our sublattices correspond to even and odd sites, then operators of the form $f^\dagger_{j,\sigma'} f_{j+1,\sigma}$ are allowed, but operators of the form $f^\dagger_{j,\uparrow}f_{j,\downarrow}$ are not. 
Nevertheless, we emphasize again that we do not require that the system is defined on a regular lattice at all.

The Hamiltonian $H_F$ in \eqref{eq:hamiltonian} preserves the fermionic particle number $\hat N = \sum_i f^\dagger_i f_i$ and therefore its energy eigenstates $\ket{E_s}$ can be chosen as eigenstates of $\hat N$. 
Number conservation is, in fact, not strictly necessary. By utilizing the well-known representation of Fermions in terms of Majorana operators \cite{Vidal2003}
\begin{align}\label{eq:majorana}
	\omega_{2j-1} := f_j^\dagger + f_j,\quad \omega_{2j} := -\mathrm i (f_j^\dagger - f_j),
\end{align}
we can show that our results also transfer to quadratic Hamiltonians involving \emph{pairing terms} of the form $f_i f_j$ or $f_i^\dagger f_j^\dagger$.
More generally, we consider arbitrary Hamiltonians quadratic in $2N$ Majorana operators of the form
\begin{align}
	H_M = \sum_{\alpha,\beta=1}^{2N} \frac{i}{4} \omega_\alpha (h_M)_{\alpha,\beta} \omega_\beta,
\end{align}
where $h_M$ is a  real and anti-symmetric $2N\times 2N$-matrix. The sublattice symmetry in this case 
means that we can partition the $2N$ Majorana operators into two subsets $A$ and $B$ with $|A|$ and $|B|$ even and such that
$(h_M)_{\alpha,\beta} = 0$ unless either $\alpha\in A$ and $\beta \in B$ or vice-versa.
The subsystem associated to $A$ (or $B$) then corresponds to the fermionic modes constructed from the Majorana operators in this set by inverting \eqref{eq:majorana}.
Without loss of generality we always assume $|A|\geq |B|$.

The energy eigenstates of Hamiltonians of the form $H_M$ and $H_F$ are \emph{Gaussian}, meaning that they fulfill Wick's theorem and therefore are completely determined by their 2-point correlation functions. We will denote them by $\ket{E_s}$ with $s\in \{0,1\}^N$  (the Fock-space of $N$ fermionic modes has dimension $2^N$).

\section{Main Result} 
To state our main result, we define $S(B|E_s)$ to be the entanglement entropy of the fermionic modes in set $B$ when the total system is in the energy eigenstate $\ket{E_s}$. We measure entropy in units of $\log(2)$. 
\begin{theorem}\label{thm:main}
If $H_M$ has a sublattice symmetry between $A$ and $B$ with $|A|\geq |B|$, then:
	\begin{enumerate}
		\item If $[H_M,\hat N]=0$, there exists an orthonormal basis of pure Gaussian energy eigenstates $\ket{E_s}$ such that
			\begin{align}
				S(B|E_s) = n(s) \in \Z_+,
			\end{align}
			for all $s=\{0,1\}^N$. At least one ground state fulfills $n(s)=|B|$. Moreover, the states $\ket{E_s}$ are simultaneous eigenstates of $\hat N$.
		\item There always exists an orthonormal basis of Gaussian energy eigenstates such that
	\begin{align}
	S(B|E_s) = |B|	
	\end{align}
			\emph{for all} $s\in\{0,1\}^N$: Every energy eigenstate is maximally entangled.
	\end{enumerate}
\end{theorem}

The theorem shows that any Hamiltonian $H_F$ of the form ~\eqref{eq:hamiltonian} has a ground state $\ket{E_{\mathrm{GS}}}$ that is maximally entangled between the two sublattices
and the entanglement in an eigenbasis of number-conserving eigenstates comes in discrete units of \emph{singlets}.
Fig.~\ref{fig:spectrum} exemplifies this quantization of eigenstate entanglement for a randomly chosen Hamiltonian with sublattice symmetry. 

The theorem also says that for a general Hamiltonian that does not preserve the fermionic particle number, \emph{all} energy eigenstates are maximally entangled between $A$ and $B$. 
In particular, this implies that also in the case of number conservation there exists a full basis of pure Gaussian energy eigenstates that are maximally entangled. 
This is compatible with our previous discussion and Fig.~\ref{fig:spectrum}, because now the energy eigenstates are not simultaneously also eigenstates of the number operator. 
The required degeneracy of the energy eigenspaces is implied by the sublattice symmetry. 

The fact that the ground state is maximally entangled between the two sublattices implies that the different fermionic modes \emph{within} the smaller sublattice $B$ are \emph{completely uncorrelated}.
Hence, all 2-point correlation functions vanish:
\begin{align}\label{eq:2pt}
	\bra{E_\mathrm{GS}} f_i^\dagger f_j\ket{E_\mathrm{GS}} = 0,\quad \forall i\neq j \in B
\end{align}
and all connected correlation functions of physical observables within $B$ vanish:
\begin{align}
	\bra{E_\mathrm{GS}} XY\! \ket{E_\mathrm{GS}} -  \bra{E_\mathrm{GS}}\! X\!\! \proj{E_{\mathrm{GS}}} Y\!\! \ket{E_\mathrm{GS}}= 0,\nonumber
\end{align}
for any observables $X$ and $Y$ supported on different $B$-modes. 

\begin{figure}[t]
	\includegraphics[width=.6\linewidth]{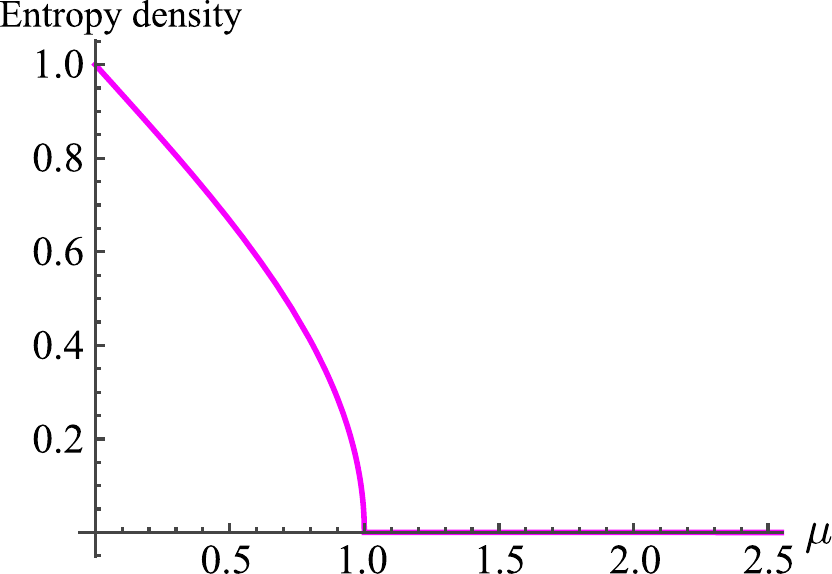}
	\caption{Entanglement entropy density of even sites in the thermodynamic limit in the ground state of the Hamiltonian $H^{XX}_F + \mu \hat N$ as a function of the chemical potential $\mu$.}
	\label{fig:qpt}
\end{figure}

\subsection{Chemical potentials.} The addition of a uniform chemical potential of the form $\mu N$  to $H_F$ {(which breaks the sublattice symmetry)} does not change the number-conserving eigenstates, but only which one of them is the ground state. 
For example, if we add a chemical potential to the free fermionic hopping chain $H_F^{XX}$, then the single-particle energies (dispersion relation) fulfill $\eps_k = \cos(2\pi k/N)+\mu$. 
If we imagine starting with $\mu=0$ and then increase $\mu$, the entanglement in the ground state decreases in unit-steps until we get the unentangled vaccum for $\mu\geq 1$.
For $\mu \geq 0$ and in the thermodynamic limit, the entropy \emph{density} of sublattice $B$ in the ground state then takes the form (see Fig.~\ref{fig:qpt}):
\begin{align*}
	\lim_{N\rightarrow \infty} \frac{S_\alpha(B|E_{\mathrm{GS}})}{|B|}  = 
		\begin{cases}
			2 - \frac{2}{\pi}\arccos(-\mu)\ &\mathrm{if} \quad \mu \leq 1,\\
			0 &\mathrm{if} \quad \mu > 1,
		\end{cases}
\end{align*}
which follows directly from the form of the single-particle energies. This matches the observation that sublattice entanglement can be used to signal quantum phase transitions \cite{Chen2006,Chen2006a,He2017}.

\subsection{Dual spin models} Often, spin-1/2 models are solved by a mapping to non-interacting fermions, see Refs.~\cite{Chapman2020,Elman2021} for the general theory of when this is possible.
For example, it is well known that the free fermionic chain with only nearest neighbour couplings $H_F^{XX}$ is dual to the XX model in terms of spins via a Jordan-Wigner transformation. 
For every \emph{connected} subsystem, the spectrum (and hence all entropies) of the reduced density matrix of the ground state of the spin model agrees with that of the corresponding reduced state of the fermionic model.
However, since the Jordan-Wigner transformation is non-local, the same is not true for \emph{disconnected} subsystems and correspondingly, the groundstate of the critical XX model is \emph{not} maximally entangled \cite{Igloi2010,Alba2010}.
{We note that the property that not all entanglement entropies match when mapping a fermionic system to a spin system is inherent to any such duality transformation, as it has to preserve the canonical anti-commutation relations and hence cannot be local.}
{To understand the consequences of the quantized entanglement in the fermionic model for the dual spin model constitute an intersting research question for future work.}

\subsection{Bosonic systems} One may be tempted to also study the case of non-interacting bosons with sublattice symmetry. However, since single-particle Hamiltonians with sublattice symmetry always have eigenmodes with negative energies, second-quantizing such Hamiltonians in a bosonic manner leads to unphysical Hamiltonians that are not bounded below. It is therefore unclear to us how to arrive at a sensible notion of sublattice symmetry for bosonic systems.

\section{Interactions}  
\begin{figure}[tb]
	\includegraphics[width=.6\linewidth]{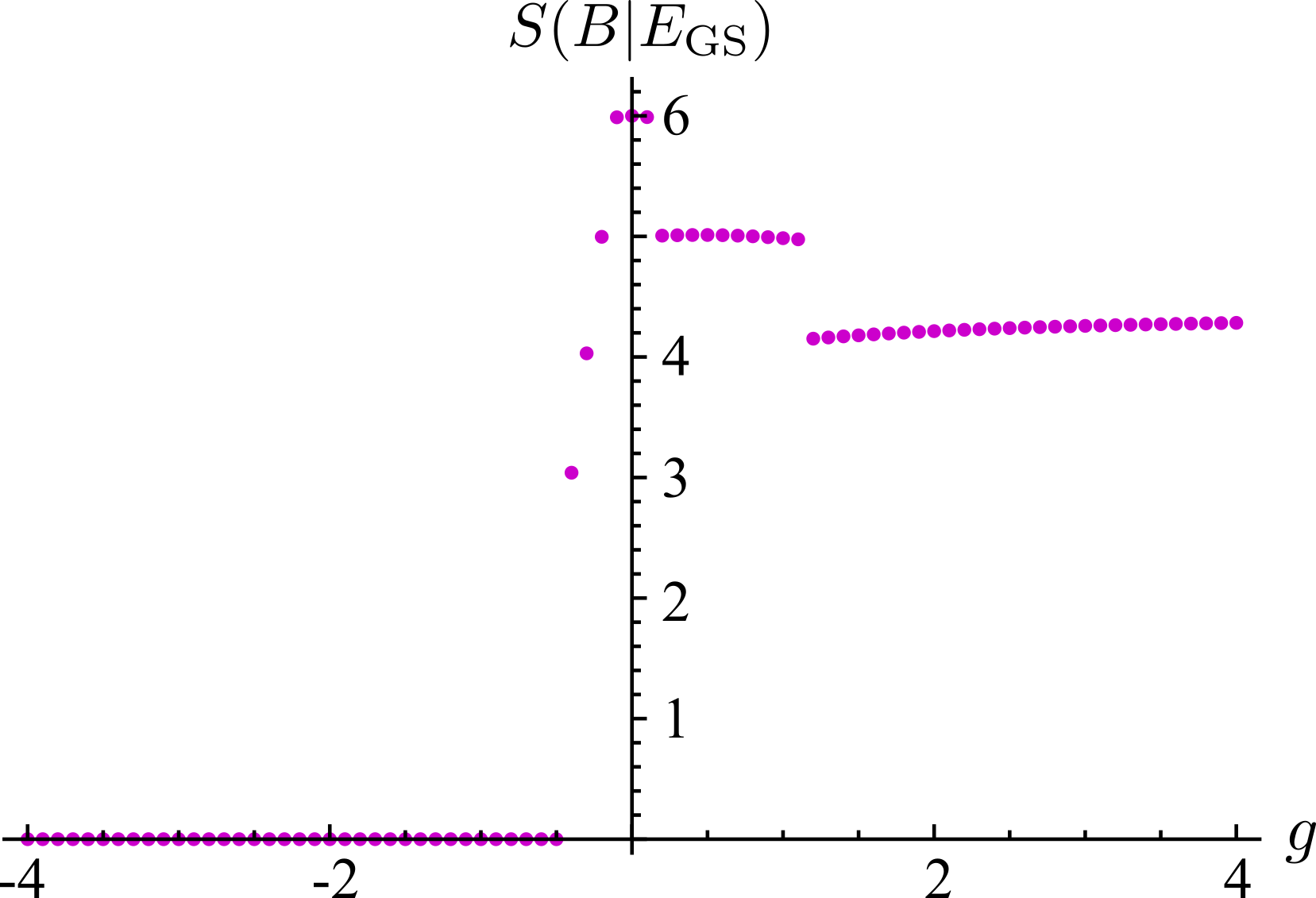}
	\caption{{The sublattice entanglement entropy $S(B|E_{\mathrm{GS}})$ as a function of the interaction strength $g$ in the interacting model $H_{\mathrm{int}}$ defined in \eqref{eq:int} for system size $N=12$.}}
	\label{fig:interactions}
\end{figure}

It is natural to wonder what happens to the sublattice entanglement once we introduce interactions. 
We expect that the quantized sublattice entanglement can only be stable to interactions, if the interaction in some sense respects the sublattice symmetry.
(While one can define chiral or particle-hole symmetries for interacting Hamiltonians, see e.g. Ref.~\cite{Zirnbauer2021}, we can only expect interesting effects for entanglement if the symmetry has a relation
to real-space.)
We therefore now consider adding a nearest-neighbor interaction to the 1d-hopping model $H^{XX}_F$ with open boundary conditions:
\begin{align}
	H_{\mathrm{int}} := H^{XX}_F + g \sum_j n_j n_{j+1},  \label{eq:int}
\end{align}
with $n_j = f_j^\dagger f_j$. In Fig.~\ref{fig:interactions} we plot the numerically extracted (via exact diagonalization) sublattice entanglement entropy as a function of the interaction strength $g$ for a lattice size $N=12$.
(Due to the above-mentioned issues with the Jordan-Wigner transformation, care must be taken in extracting the entanglement entropy numerically.)
Interestingly, while strong interactions destroy the \emph{maximum} entanglement in the ground state, we see that the entanglement remains almost quantized in terms of singlets and finite even for strong repulsive ($g>0$) interactions. 
For attractive interactions ($g<0$), the entanglement also drops almost quantized units of singlets, but relatively quickly to zero. 
The latter effect can be heuristically explained by the fact that strongly attractive interactions favor a completely filled system, which has zero sublattice entanglement. 
The striking quantization effect for repulsive interactions, on the other hand, requires further investigation.

\section{Proof of the number-conserving case} Since we believe that the proof of the number-conserving case is physically more illuminating, we only give this proof in the main text. The proof of the full Theorem~\ref{thm:main}, which relies on general properties of quasi-free fermionic systems in the Majorana representation, is given in the Appendix.  

Due to the sublattice symmetry, we may write the single-particle Hamiltonian $h$ in \eqref{eq:hamiltonian} as
\begin{align}\label{eq:sublattice}
	h = \begin{pmatrix} 0 & h_{AB}\\
		h_{AB}^\dagger & 0
		\end{pmatrix},
\end{align}
where $h_{AB}$ is a complex $|A|\times |B|$-matrix.
The Hamiltonian $H_F$ may be diagonalized by first obtaining an orthonormal basis of eigenvectors $\psi_k$ with eigenvalues $\eps_k$ of $h$ and then introducing the normal modes 
\begin{align}
	a^\dagger_k = \sum_i \psi_k(i) f_i^\dagger.
\end{align}
Then $H_F$ takes the form
\begin{align}
	H_F = \sum_{k=1}^{N} \eps_k a^\dagger_k a_k
\end{align}
and energy eigenstates take the form
\begin{align}\label{eq:eigenstate}
	\ket{E_s} = (a^\dagger_{1})^{s_1} \cdots (a^\dagger_N)^{s_N}\ket{0},\quad s_i\in \{0,1\},
\end{align}
with energies $E_s = \sum_{k=1}^{N} s_k \eps_k$ and $\ket{0}$ denoting the fermionic vaccum annihilated by the operators $f_j$.
The sublattice symmetry has strong implications for the structure of the eigenvectors $\psi_k$ of $h$,
which will be crucial for our results. 
This structure is summarized in the following Lemma, whose proof is given in the Appendix.
In essence it says that the eigenfunctions $\psi_k$ of $h$ can be chosen in such a way that they 
are either evenly distributed between $A$ and $B$ or completely supported on $A$ (this is where the assumption $|A|\geq |B|$ enters).
\begin{lemma}\label{lemma:basic}
Let $h$ be a Hermitian $N\times N$-matrix of the form \eqref{eq:sublattice}.
	Then there exists an orthonormal basis of vectors $\{\psi_{B,k}\}_{k=1}^{|B|}$ of $\mathbb C^{|B|}$, an isometry $V:\mathbb C^{|B|}\rightarrow \mathbb C^{|A|}$
and $|A|-|B|$ orthonormal vectors $\phi_l \in \mathbb C^{|A|}$ such that the vectors
	\begin{align}
	\psi_k = \begin{cases}
			\frac{1}{\sqrt 2}(V\psi_{B,k},\psi_{B,k})^\top\quad &k=1,\ldots,|B|\\
			\frac{1}{\sqrt 2}(V\psi_{B,k},-\psi_{B,})^\top\quad &k=|B|+1,\ldots, 2|B|\\
			(\phi_{k-2|B|},0)^\top			 \quad &k=2|B|+1,\ldots, N,
		\end{cases}\nonumber
	\end{align}
provide an orthonormal basis of eigenvectors of $h$ with $h\psi_k = 0$ for $k=2|B|+1,\ldots,N$.
\end{lemma}
Due to the sublattice symmetry it follows immediately that if $(\psi_{A,k},\psi_{B,k})^\top$ is an eigenvector of $h$ with eigenvalue $\eps_k$, 
then $(\psi_{A,k},-\psi_{B,k})^\top$ is an eigenvector with eigenvalue $-\eps_k$.
Hence, according to the labelling introduced by the Lemma, we have $\eps_{k+|B|} = -\eps_k$ for $k=1,\ldots,|B|$.
We will say that the Hamiltonian is \emph{regular} if the only eigenvectors with $\eps_k=0$, so-called \emph{zero-modes}, are those with $k=2|B|+1,\ldots,N$. We will call these zero-modes $A$-modes, since they are completely supported on $A$. 
If, despite the $A$-modes, there are $m_z$ additional zero-modes, we will call them \emph{spurious} zero-modes.

\subsection{Entanglement in the energy eigenstates.}
We will now show that the entanglement in the energy eigenstate $\ket{E_s}$ consists of 
\begin{align}
	n(s) := \sum_{k=1}^{|B|} (s_k \oplus s_{k+|B|})\in \Z_+
\end{align}
many shared \emph{singlets} between $A$ and $B$. Here, $\oplus$ denotes addition $\mod 2$.
Therefore, all \emph{R\'enyi entropies} fulfill
\begin{align}\label{eq:entanglement}
	S_\alpha(B|E_s) := S_\alpha(\Tr_A[\proj{E_s}]) = n(s),
\end{align}
independent of the parameter $\alpha$. 
The entanglement entropies are therefore quantized and the ground states consists of at least $n(s)=|B|-m_z$ singlets. 
In particular,  if there are no spurious zero-modes, all ground states are maximally entangled. Even in the presence of spurious zero-modes, we can choose at least one ground state that is maximally entangled.

We now make use of the normal-mode decomposition provided by Lemma to show \eqref{eq:entanglement}. To do that we write, for $1\leq k\leq |B|$,
\begin{align}
	a_k^\dagger  
	&= \frac{1}{\sqrt{2}}\sum_{j\in A} (V\psi_{B,k})(j) f_j^\dagger + \frac{1}{\sqrt{2}}\sum_{j\in B}\psi_{B,k}(j)f_j^\dagger \\
	&=: \frac{1}{\sqrt{2}} a_{A,k}^\dagger + \frac{1}{\sqrt{2}} a_{B,k}^\dagger.
\end{align}
A similar calculation for $a_{k+|B|}^\dagger$ shows
\begin{align}
	a_{A,k+|B|}^\dagger = a_{A,k}^\dagger,\quad a_{B,k+|B|}^\dagger = - a_{B,k}^\dagger.
\end{align}
The $a_{B,k}^\dagger$ with $k=1,\ldots,|B|$ provide a complete set of fermionic creation operators on $B$.
Similarly, the $a_{A,k}^\dagger$ with $k=1,\ldots,|B|$ together with the $a_k^\dagger$ for $k=2|B|+1,\ldots,N$ provide a complete set of fermionic creation operators for $A$.
Thus, each of the subsystems $A$ and $B$ factorizes into the normal modes $a_{A,k}^\dagger$ and $a_{B,k}^\dagger$, respectively. 
Therefore, we can write the fermionic vacuum on $A$ as \footnote{We are aware that strictly speaking there is no tensor-product structure. But there is no harm in writing it like this.}
\begin{align}
	\ket{0}_A = \otimes_{k=1}^{|B|} \ket{0}_{k,A} \otimes_{l=1}^{|A|-|B|}\ket{0}_{l,A},
\end{align}
where $\ket{0}_{l,A}$ denotes the vacuum associated to the zero-mode $\phi_l$ on $A$.
Similarly, $\ket{0}_B = \otimes_{k=1}^{|B|} \ket{0}_{k,B}$. 
We will also write $a_{A,k}^\dagger \ket{0}_A = \ket{1_k}_A$ and similarly for $B$.
Consequently, we have for $k=1,\ldots,M$:
\begin{align}\label{eq:singlet1}
	a_k^\dagger\! \ket{0} 
	&=\frac{1}{\sqrt 2}\left[ \ket{1_k}_A\! \ket{0}_{B} + \ket{0}_{A}\! \ket{1_k}_{B}\right] =: \ket{\Phi^+_k},\\
	a_{k+M}^\dagger\!\ket{0} &= \frac{1}{\sqrt 2}\left[ \ket{1_k}_A\! \ket{0}_{B} - \ket{0}_{A}\! \ket{1_k}_{B}\right] =: \ket{\Phi^-_k},\nonumber
\end{align}
which are singlets, but
\begin{align}\label{eq:product}
	a_{k+M}^\dagger a_k^\dagger\ket{0} = \ket{1_k}_A \ket{1_k}_{B},
\end{align}
which is again a product-state. 
According to \eqref{eq:eigenstate}, the energy eigenstate $\ket{E_s}$ is constructed by filling up the normal modes for which $s_k=1$.
From \eqref{eq:singlet1} we see that each such filled normal mode with $k=1,\ldots,2|B|$ adds one singlet between $A$ and $B$, but additionally filling up the normal mode with the same mode-number $\mod |B|$ removes the singlet again due to \eqref{eq:product}. 
Since the $A$-modes $a_k^\dagger$ with $k=2|B|+1,\ldots,N$ are completely supported on $A$, the entanglement between $A$ and $B$ is not affected by them being filled or not. Thus each energy eigenstate consists of exactly $n(s)$ singlets between $A$ and $B$.

\section{Conclusions}
We have shown that sublattice symmetry of a Hamiltonian of non-interacting fermions implies that {there always exists a basis of energy eigenstates such that} the entanglement between the two sublattices is quantized in terms of singlets for all energy eigenstates.
Moreover, there always exists a ground state for which the entanglement is maximal.

The sublattice symmetry we considered implies the existence of an (orthogonal) matrix $S$ such that $S^2=\1$ and
\begin{align}\label{eq:chiral}
ShS = -h,
\end{align}
placing our considered systems in the chiral symmetry classes of the \emph{tenfold way} \cite{Altland1997,Heinzner2005,Kitaev2009,Ryu2010,Ludwig2015}.
In contrast to the classification in the tenfold way, we here require the interpretation of the sublattice symmetry in terms of real space, since we are dealing with spatial entanglement.
Our result shows that every non-trivial single-particle Hamiltonian $h$ fulfilling \eqref{eq:chiral} has a ground state that is maximally entangled when second-quantized. 
In particular, this means that no such Hamiltonian can have an unentangled state (between $A$ and $B$) as unique ground state.  
The maximum entanglement in at least one ground state is thus an invariant of all the corresponding second-quantized Hamiltonians and does not distinguish different topological phases.

Abstractly, chiral symmetry is defined as the composition of time-reversal and particle-hole symmetry.  
It may be interesting to study the effects on the entanglement structure of energy eigenstates that time-reversal and particle-hole symmetries have individually, although here the relationship to real space is less clear.

{As already mentioned above, it is currently unclear what consequences the quantized entanglement of free fermionic systems has for dual spin systems. We believe this to be an interesting open question. 

Finally, we presented preliminary results indicating that the quantization of sublattice entanglement can extend to the case of interacting fermionic systems. Thoroughly understanding the effect
of sublattice symmetry for the entanglement in interacting fermionic systems presents an interesting challenge for future work.} 

\begin{acknowledgements}
We thank Zolt\'an Zimbor\'as and Christoph Karrasch for valuable correspondence.
Support by the DFG through SFB 1227 (DQ-mat), Quantum Valley Lower Saxony, and funding
 by the Deutsche Forschungsgemeinschaft (DFG, German Research Foundation) under Germanys Excellence Strategy EXC-2123 QuantumFrontiers 390837967 is also
 acknowledged.
\end{acknowledgements}

\bibliographystyle{apsrev4-1}

\begin{thebibliography}{0}%
\makeatletter
\providecommand \@ifxundefined [1]{%
 \@ifx{#1\undefined}
}%
\providecommand \@ifnum [1]{%
 \ifnum #1\expandafter \@firstoftwo
 \else \expandafter \@secondoftwo
 \fi
}%
\providecommand \@ifx [1]{%
 \ifx #1\expandafter \@firstoftwo
 \else \expandafter \@secondoftwo
 \fi
}%
\providecommand \natexlab [1]{#1}%
\providecommand \enquote  [1]{``#1''}%
\providecommand \bibnamefont  [1]{#1}%
\providecommand \bibfnamefont [1]{#1}%
\providecommand \citenamefont [1]{#1}%
\providecommand \href@noop [0]{\@secondoftwo}%
\providecommand \href [0]{\begingroup \@sanitize@url \@href}%
\providecommand \@href[1]{\@@startlink{#1}\@@href}%
\providecommand \@@href[1]{\endgroup#1\@@endlink}%
\providecommand \@sanitize@url [0]{\catcode `\\12\catcode `\$12\catcode
  `\&12\catcode `\#12\catcode `\^12\catcode `\_12\catcode `\%12\relax}%
\providecommand \@@startlink[1]{}%
\providecommand \@@endlink[0]{}%
\providecommand \url  [0]{\begingroup\@sanitize@url \@url }%
\providecommand \@url [1]{\endgroup\@href {#1}{\urlprefix }}%
\providecommand \urlprefix  [0]{URL }%
\providecommand \Eprint [0]{\href }%
\providecommand \doibase [0]{http://dx.doi.org/}%
\providecommand \selectlanguage [0]{\@gobble}%
\providecommand \bibinfo  [0]{\@secondoftwo}%
\providecommand \bibfield  [0]{\@secondoftwo}%
\providecommand \translation [1]{[#1]}%
\providecommand \BibitemOpen [0]{}%
\providecommand \bibitemStop [0]{}%
\providecommand \bibitemNoStop [0]{.\EOS\space}%
\providecommand \EOS [0]{\spacefactor3000\relax}%
\providecommand \BibitemShut  [1]{\csname bibitem#1\endcsname}%
\let\auto@bib@innerbib\@empty
\end{thebibliography}%


\begin{thebibliography}{46}%
\makeatletter
\providecommand \@ifxundefined [1]{%
 \@ifx{#1\undefined}
}%
\providecommand \@ifnum [1]{%
 \ifnum #1\expandafter \@firstoftwo
 \else \expandafter \@secondoftwo
 \fi
}%
\providecommand \@ifx [1]{%
 \ifx #1\expandafter \@firstoftwo
 \else \expandafter \@secondoftwo
 \fi
}%
\providecommand \natexlab [1]{#1}%
\providecommand \enquote  [1]{``#1''}%
\providecommand \bibnamefont  [1]{#1}%
\providecommand \bibfnamefont [1]{#1}%
\providecommand \citenamefont [1]{#1}%
\providecommand \href@noop [0]{\@secondoftwo}%
\providecommand \href [0]{\begingroup \@sanitize@url \@href}%
\providecommand \@href[1]{\@@startlink{#1}\@@href}%
\providecommand \@@href[1]{\endgroup#1\@@endlink}%
\providecommand \@sanitize@url [0]{\catcode `\\12\catcode `\$12\catcode
  `\&12\catcode `\#12\catcode `\^12\catcode `\_12\catcode `\%12\relax}%
\providecommand \@@startlink[1]{}%
\providecommand \@@endlink[0]{}%
\providecommand \url  [0]{\begingroup\@sanitize@url \@url }%
\providecommand \@url [1]{\endgroup\@href {#1}{\urlprefix }}%
\providecommand \urlprefix  [0]{URL }%
\providecommand \Eprint [0]{\href }%
\providecommand \doibase [0]{http://dx.doi.org/}%
\providecommand \selectlanguage [0]{\@gobble}%
\providecommand \bibinfo  [0]{\@secondoftwo}%
\providecommand \bibfield  [0]{\@secondoftwo}%
\providecommand \translation [1]{[#1]}%
\providecommand \BibitemOpen [0]{}%
\providecommand \bibitemStop [0]{}%
\providecommand \bibitemNoStop [0]{.\EOS\space}%
\providecommand \EOS [0]{\spacefactor3000\relax}%
\providecommand \BibitemShut  [1]{\csname bibitem#1\endcsname}%
\let\auto@bib@innerbib\@empty
\bibitem [{\citenamefont {Eisert}\ \emph {et~al.}(2010)\citenamefont {Eisert},
  \citenamefont {Cramer},\ and\ \citenamefont {Plenio}}]{Eisert2010}%
  \BibitemOpen
  \bibfield  {author} {\bibinfo {author} {\bibfnamefont {J.}~\bibnamefont
  {Eisert}}, \bibinfo {author} {\bibfnamefont {M.}~\bibnamefont {Cramer}}, \
  and\ \bibinfo {author} {\bibfnamefont {M.~B.}\ \bibnamefont {Plenio}},\
  }\href {\doibase 10.1103/RevModPhys.82.277} {\bibfield  {journal} {\bibinfo
  {journal} {Rev. Mod. Phys.}\ }\textbf {\bibinfo {volume} {82}},\ \bibinfo
  {pages} {277} (\bibinfo {year} {2010})}\BibitemShut {NoStop}%
\bibitem [{\citenamefont {Schollwöck}(2011)}]{Schollwoeck2011}%
  \BibitemOpen
  \bibfield  {author} {\bibinfo {author} {\bibfnamefont {U.}~\bibnamefont
  {Schollwöck}},\ }\href {\doibase 10.1016/j.aop.2010.09.012} {\bibfield
  {journal} {\bibinfo  {journal} {Annals of Physics}\ }\textbf {\bibinfo
  {volume} {326}},\ \bibinfo {pages} {96} (\bibinfo {year} {2011})}\BibitemShut
  {NoStop}%
\bibitem [{\citenamefont {Cirac}\ \emph {et~al.}(2021)\citenamefont {Cirac},
  \citenamefont {P{\'{e}}rez-Garc{\'{\i}}a}, \citenamefont {Schuch},\ and\
  \citenamefont {Verstraete}}]{Cirac2021}%
  \BibitemOpen
  \bibfield  {author} {\bibinfo {author} {\bibfnamefont {J.~I.}\ \bibnamefont
  {Cirac}}, \bibinfo {author} {\bibfnamefont {D.}~\bibnamefont
  {P{\'{e}}rez-Garc{\'{\i}}a}}, \bibinfo {author} {\bibfnamefont
  {N.}~\bibnamefont {Schuch}}, \ and\ \bibinfo {author} {\bibfnamefont
  {F.}~\bibnamefont {Verstraete}},\ }\href {\doibase
  10.1103/revmodphys.93.045003} {\bibfield  {journal} {\bibinfo  {journal}
  {Reviews of Modern Physics}\ }\textbf {\bibinfo {volume} {93}},\ \bibinfo
  {pages} {045003} (\bibinfo {year} {2021})}\BibitemShut {NoStop}%
\bibitem [{\citenamefont {Wen}(1990)}]{Wen1990}%
  \BibitemOpen
  \bibfield  {author} {\bibinfo {author} {\bibfnamefont {X.-G.}\ \bibnamefont
  {Wen}},\ }\href {\doibase 10.1142/s0217979290000139} {\bibfield  {journal}
  {\bibinfo  {journal} {International Journal of Modern Physics B}\ }\textbf
  {\bibinfo {volume} {04}},\ \bibinfo {pages} {239} (\bibinfo {year}
  {1990})}\BibitemShut {NoStop}%
\bibitem [{\citenamefont {Kitaev}(2003)}]{Kitaev2003}%
  \BibitemOpen
  \bibfield  {author} {\bibinfo {author} {\bibfnamefont {A.}~\bibnamefont
  {Kitaev}},\ }\href {\doibase 10.1016/s0003-4916(02)00018-0} {\bibfield
  {journal} {\bibinfo  {journal} {Annals of Physics}\ }\textbf {\bibinfo
  {volume} {303}},\ \bibinfo {pages} {2} (\bibinfo {year} {2003})}\BibitemShut
  {NoStop}%
\bibitem [{\citenamefont {Kitaev}(2006)}]{Kitaev2006}%
  \BibitemOpen
  \bibfield  {author} {\bibinfo {author} {\bibfnamefont {A.}~\bibnamefont
  {Kitaev}},\ }\href {\doibase 10.1016/j.aop.2005.10.005} {\bibfield  {journal}
  {\bibinfo  {journal} {Annals of Physics}\ }\textbf {\bibinfo {volume}
  {321}},\ \bibinfo {pages} {2} (\bibinfo {year} {2006})}\BibitemShut {NoStop}%
\bibitem [{\citenamefont {Wen}(2013)}]{Wen2013}%
  \BibitemOpen
  \bibfield  {author} {\bibinfo {author} {\bibfnamefont {X.-G.}\ \bibnamefont
  {Wen}},\ }\href {\doibase 10.1155/2013/198710} {\bibfield  {journal}
  {\bibinfo  {journal} {{ISRN} Condensed Matter Physics}\ }\textbf {\bibinfo
  {volume} {2013}},\ \bibinfo {pages} {1} (\bibinfo {year} {2013})}\BibitemShut
  {NoStop}%
\bibitem [{\citenamefont {Deutsch}(1991)}]{Deutsch1991}%
  \BibitemOpen
  \bibfield  {author} {\bibinfo {author} {\bibfnamefont {J.~M.}\ \bibnamefont
  {Deutsch}},\ }\href {\doibase 10.1103/PhysRevA.43.2046} {\bibfield  {journal}
  {\bibinfo  {journal} {Phys. Rev. A}\ }\textbf {\bibinfo {volume} {43}},\
  \bibinfo {pages} {2046} (\bibinfo {year} {1991})}\BibitemShut {NoStop}%
\bibitem [{\citenamefont {Srednicki}(1994)}]{Srednicki1994}%
  \BibitemOpen
  \bibfield  {author} {\bibinfo {author} {\bibfnamefont {M.}~\bibnamefont
  {Srednicki}},\ }\href {\doibase 10.1103/PhysRevE.50.888} {\bibfield
  {journal} {\bibinfo  {journal} {Phys. Rev. E}\ }\textbf {\bibinfo {volume}
  {50}},\ \bibinfo {pages} {888} (\bibinfo {year} {1994})}\BibitemShut
  {NoStop}%
\bibitem [{\citenamefont {D'Alessio}\ \emph {et~al.}(2016)\citenamefont
  {D'Alessio}, \citenamefont {Kafri}, \citenamefont {Polkovnikov},\ and\
  \citenamefont {Rigol}}]{DAlessio2016}%
  \BibitemOpen
  \bibfield  {author} {\bibinfo {author} {\bibfnamefont {L.}~\bibnamefont
  {D'Alessio}}, \bibinfo {author} {\bibfnamefont {Y.}~\bibnamefont {Kafri}},
  \bibinfo {author} {\bibfnamefont {A.}~\bibnamefont {Polkovnikov}}, \ and\
  \bibinfo {author} {\bibfnamefont {M.}~\bibnamefont {Rigol}},\ }\href
  {\doibase 10.1080/00018732.2016.1198134} {\bibfield  {journal} {\bibinfo
  {journal} {Adv. Phys.}\ }\textbf {\bibinfo {volume} {65}},\ \bibinfo {pages}
  {239} (\bibinfo {year} {2016})}\BibitemShut {NoStop}%
\bibitem [{\citenamefont {Gogolin}\ and\ \citenamefont
  {Eisert}(2016)}]{Gogolin2016}%
  \BibitemOpen
  \bibfield  {author} {\bibinfo {author} {\bibfnamefont {C.}~\bibnamefont
  {Gogolin}}\ and\ \bibinfo {author} {\bibfnamefont {J.}~\bibnamefont
  {Eisert}},\ }\href {\doibase 10.1088/0034-4885/79/5/056001} {\bibfield
  {journal} {\bibinfo  {journal} {Rep. Prog. Phys.}\ }\textbf {\bibinfo
  {volume} {79}},\ \bibinfo {pages} {56001} (\bibinfo {year} {2016})},\ \Eprint
  {http://arxiv.org/abs/1503.07538} {arXiv:1503.07538} \BibitemShut {NoStop}%
\bibitem [{\citenamefont {Gornyi}\ \emph {et~al.}(2005)\citenamefont {Gornyi},
  \citenamefont {Mirlin},\ and\ \citenamefont {Polyakov}}]{Gornyi2005}%
  \BibitemOpen
  \bibfield  {author} {\bibinfo {author} {\bibfnamefont {I.~V.}\ \bibnamefont
  {Gornyi}}, \bibinfo {author} {\bibfnamefont {A.~D.}\ \bibnamefont {Mirlin}},
  \ and\ \bibinfo {author} {\bibfnamefont {D.~G.}\ \bibnamefont {Polyakov}},\
  }\href {\doibase 10.1103/physrevlett.95.206603} {\bibfield  {journal}
  {\bibinfo  {journal} {Physical Review Letters}\ }\textbf {\bibinfo {volume}
  {95}},\ \bibinfo {pages} {206603} (\bibinfo {year} {2005})}\BibitemShut
  {NoStop}%
\bibitem [{\citenamefont {Basko}\ \emph {et~al.}(2006)\citenamefont {Basko},
  \citenamefont {Aleiner},\ and\ \citenamefont {Altshuler}}]{Basko2006}%
  \BibitemOpen
  \bibfield  {author} {\bibinfo {author} {\bibfnamefont {D.}~\bibnamefont
  {Basko}}, \bibinfo {author} {\bibfnamefont {I.}~\bibnamefont {Aleiner}}, \
  and\ \bibinfo {author} {\bibfnamefont {B.}~\bibnamefont {Altshuler}},\ }\href
  {\doibase 10.1016/j.aop.2005.11.014} {\bibfield  {journal} {\bibinfo
  {journal} {Annals of Physics}\ }\textbf {\bibinfo {volume} {321}},\ \bibinfo
  {pages} {1126} (\bibinfo {year} {2006})}\BibitemShut {NoStop}%
\bibitem [{\citenamefont {Nandkishore}\ and\ \citenamefont
  {Huse}(2015)}]{Nandkishore2015}%
  \BibitemOpen
  \bibfield  {author} {\bibinfo {author} {\bibfnamefont {R.}~\bibnamefont
  {Nandkishore}}\ and\ \bibinfo {author} {\bibfnamefont {D.~A.}\ \bibnamefont
  {Huse}},\ }\href {\doibase 10.1146/annurev-conmatphys-031214-014726}
  {\bibfield  {journal} {\bibinfo  {journal} {Ann. Rev. Cond. Matter Phys.}\
  }\textbf {\bibinfo {volume} {6}},\ \bibinfo {pages} {15} (\bibinfo {year}
  {2015})}\BibitemShut {NoStop}%
\bibitem [{\citenamefont {Abanin}\ \emph {et~al.}(2019)\citenamefont {Abanin},
  \citenamefont {Altman}, \citenamefont {Bloch},\ and\ \citenamefont
  {Serbyn}}]{Abanin2019}%
  \BibitemOpen
  \bibfield  {author} {\bibinfo {author} {\bibfnamefont {D.~A.}\ \bibnamefont
  {Abanin}}, \bibinfo {author} {\bibfnamefont {E.}~\bibnamefont {Altman}},
  \bibinfo {author} {\bibfnamefont {I.}~\bibnamefont {Bloch}}, \ and\ \bibinfo
  {author} {\bibfnamefont {M.}~\bibnamefont {Serbyn}},\ }\href {\doibase
  10.1103/revmodphys.91.021001} {\bibfield  {journal} {\bibinfo  {journal}
  {Reviews of Modern Physics}\ }\textbf {\bibinfo {volume} {91}},\ \bibinfo
  {pages} {021001} (\bibinfo {year} {2019})}\BibitemShut {NoStop}%
\bibitem [{\citenamefont {Bernien}\ \emph {et~al.}(2017)\citenamefont
  {Bernien}, \citenamefont {Schwartz}, \citenamefont {Keesling}, \citenamefont
  {Levine}, \citenamefont {Omran}, \citenamefont {Pichler}, \citenamefont
  {Choi}, \citenamefont {Zibrov}, \citenamefont {Endres}, \citenamefont
  {Greiner}, \citenamefont {Vuleti{\'{c}}},\ and\ \citenamefont
  {Lukin}}]{Bernien2017}%
  \BibitemOpen
  \bibfield  {author} {\bibinfo {author} {\bibfnamefont {H.}~\bibnamefont
  {Bernien}}, \bibinfo {author} {\bibfnamefont {S.}~\bibnamefont {Schwartz}},
  \bibinfo {author} {\bibfnamefont {A.}~\bibnamefont {Keesling}}, \bibinfo
  {author} {\bibfnamefont {H.}~\bibnamefont {Levine}}, \bibinfo {author}
  {\bibfnamefont {A.}~\bibnamefont {Omran}}, \bibinfo {author} {\bibfnamefont
  {H.}~\bibnamefont {Pichler}}, \bibinfo {author} {\bibfnamefont
  {S.}~\bibnamefont {Choi}}, \bibinfo {author} {\bibfnamefont {A.~S.}\
  \bibnamefont {Zibrov}}, \bibinfo {author} {\bibfnamefont {M.}~\bibnamefont
  {Endres}}, \bibinfo {author} {\bibfnamefont {M.}~\bibnamefont {Greiner}},
  \bibinfo {author} {\bibfnamefont {V.}~\bibnamefont {Vuleti{\'{c}}}}, \ and\
  \bibinfo {author} {\bibfnamefont {M.~D.}\ \bibnamefont {Lukin}},\ }\href
  {\doibase 10.1038/nature24622} {\bibfield  {journal} {\bibinfo  {journal}
  {Nature}\ }\textbf {\bibinfo {volume} {551}},\ \bibinfo {pages} {579}
  (\bibinfo {year} {2017})}\BibitemShut {NoStop}%
\bibitem [{\citenamefont {Turner}\ \emph
  {et~al.}(2018{\natexlab{a}})\citenamefont {Turner}, \citenamefont
  {Michailidis}, \citenamefont {Abanin}, \citenamefont {Serbyn},\ and\
  \citenamefont {Papi{\'{c}}}}]{Turner2018}%
  \BibitemOpen
  \bibfield  {author} {\bibinfo {author} {\bibfnamefont {C.~J.}\ \bibnamefont
  {Turner}}, \bibinfo {author} {\bibfnamefont {A.~A.}\ \bibnamefont
  {Michailidis}}, \bibinfo {author} {\bibfnamefont {D.~A.}\ \bibnamefont
  {Abanin}}, \bibinfo {author} {\bibfnamefont {M.}~\bibnamefont {Serbyn}}, \
  and\ \bibinfo {author} {\bibfnamefont {Z.}~\bibnamefont {Papi{\'{c}}}},\
  }\href {\doibase 10.1038/s41567-018-0137-5} {\bibfield  {journal} {\bibinfo
  {journal} {Nat. Phys.}\ }\textbf {\bibinfo {volume} {14}},\ \bibinfo {pages}
  {745} (\bibinfo {year} {2018}{\natexlab{a}})}\BibitemShut {NoStop}%
\bibitem [{\citenamefont {Turner}\ \emph
  {et~al.}(2018{\natexlab{b}})\citenamefont {Turner}, \citenamefont
  {Michailidis}, \citenamefont {Abanin}, \citenamefont {Serbyn},\ and\
  \citenamefont {Papi{\'{c}}}}]{Turner2018a}%
  \BibitemOpen
  \bibfield  {author} {\bibinfo {author} {\bibfnamefont {C.~J.}\ \bibnamefont
  {Turner}}, \bibinfo {author} {\bibfnamefont {A.~A.}\ \bibnamefont
  {Michailidis}}, \bibinfo {author} {\bibfnamefont {D.~A.}\ \bibnamefont
  {Abanin}}, \bibinfo {author} {\bibfnamefont {M.}~\bibnamefont {Serbyn}}, \
  and\ \bibinfo {author} {\bibfnamefont {Z.}~\bibnamefont {Papi{\'{c}}}},\
  }\href {\doibase 10.1103/physrevb.98.155134} {\bibfield  {journal} {\bibinfo
  {journal} {Phys. Rev. B}\ }\textbf {\bibinfo {volume} {98}} (\bibinfo {year}
  {2018}{\natexlab{b}}),\ 10.1103/physrevb.98.155134}\BibitemShut {NoStop}%
\bibitem [{\citenamefont {Ho}\ \emph {et~al.}(2019)\citenamefont {Ho},
  \citenamefont {Choi}, \citenamefont {Pichler},\ and\ \citenamefont
  {Lukin}}]{Ho2019}%
  \BibitemOpen
  \bibfield  {author} {\bibinfo {author} {\bibfnamefont {W.~W.}\ \bibnamefont
  {Ho}}, \bibinfo {author} {\bibfnamefont {S.}~\bibnamefont {Choi}}, \bibinfo
  {author} {\bibfnamefont {H.}~\bibnamefont {Pichler}}, \ and\ \bibinfo
  {author} {\bibfnamefont {M.~D.}\ \bibnamefont {Lukin}},\ }\href {\doibase
  10.1103/physrevlett.122.040603} {\bibfield  {journal} {\bibinfo  {journal}
  {Physical Review Letters}\ }\textbf {\bibinfo {volume} {122}},\ \bibinfo
  {pages} {040603} (\bibinfo {year} {2019})}\BibitemShut {NoStop}%
\bibitem [{\citenamefont {Choi}\ \emph {et~al.}(2019)\citenamefont {Choi},
  \citenamefont {Turner}, \citenamefont {Pichler}, \citenamefont {Ho},
  \citenamefont {Michailidis}, \citenamefont {Papi{\'{c}}}, \citenamefont
  {Serbyn}, \citenamefont {Lukin},\ and\ \citenamefont {Abanin}}]{Choi2019}%
  \BibitemOpen
  \bibfield  {author} {\bibinfo {author} {\bibfnamefont {S.}~\bibnamefont
  {Choi}}, \bibinfo {author} {\bibfnamefont {C.~J.}\ \bibnamefont {Turner}},
  \bibinfo {author} {\bibfnamefont {H.}~\bibnamefont {Pichler}}, \bibinfo
  {author} {\bibfnamefont {W.~W.}\ \bibnamefont {Ho}}, \bibinfo {author}
  {\bibfnamefont {A.~A.}\ \bibnamefont {Michailidis}}, \bibinfo {author}
  {\bibfnamefont {Z.}~\bibnamefont {Papi{\'{c}}}}, \bibinfo {author}
  {\bibfnamefont {M.}~\bibnamefont {Serbyn}}, \bibinfo {author} {\bibfnamefont
  {M.~D.}\ \bibnamefont {Lukin}}, \ and\ \bibinfo {author} {\bibfnamefont
  {D.~A.}\ \bibnamefont {Abanin}},\ }\href {\doibase
  10.1103/physrevlett.122.220603} {\bibfield  {journal} {\bibinfo  {journal}
  {Physical Review Letters}\ }\textbf {\bibinfo {volume} {122}},\ \bibinfo
  {pages} {220603} (\bibinfo {year} {2019})}\BibitemShut {NoStop}%
\bibitem [{\citenamefont {Schecter}\ and\ \citenamefont
  {Iadecola}(2019)}]{Schecter2019}%
  \BibitemOpen
  \bibfield  {author} {\bibinfo {author} {\bibfnamefont {M.}~\bibnamefont
  {Schecter}}\ and\ \bibinfo {author} {\bibfnamefont {T.}~\bibnamefont
  {Iadecola}},\ }\href {\doibase 10.1103/physrevlett.123.147201} {\bibfield
  {journal} {\bibinfo  {journal} {Physical Review Letters}\ }\textbf {\bibinfo
  {volume} {123}} (\bibinfo {year} {2019}),\
  10.1103/physrevlett.123.147201}\BibitemShut {NoStop}%
\bibitem [{\citenamefont {Alhambra}\ \emph {et~al.}(2020)\citenamefont
  {Alhambra}, \citenamefont {Anshu},\ and\ \citenamefont
  {Wilming}}]{Alhambra2020}%
  \BibitemOpen
  \bibfield  {author} {\bibinfo {author} {\bibfnamefont {{\'{A}}.~M.}\
  \bibnamefont {Alhambra}}, \bibinfo {author} {\bibfnamefont {A.}~\bibnamefont
  {Anshu}}, \ and\ \bibinfo {author} {\bibfnamefont {H.}~\bibnamefont
  {Wilming}},\ }\href {\doibase 10.1103/physrevb.101.205107} {\bibfield
  {journal} {\bibinfo  {journal} {Physical Review B}\ }\textbf {\bibinfo
  {volume} {101}},\ \bibinfo {pages} {205107} (\bibinfo {year}
  {2020})}\BibitemShut {NoStop}%
\bibitem [{\citenamefont {Serbyn}\ \emph {et~al.}(2021)\citenamefont {Serbyn},
  \citenamefont {Abanin},\ and\ \citenamefont {Papi{\'{c}}}}]{Serbyn2021}%
  \BibitemOpen
  \bibfield  {author} {\bibinfo {author} {\bibfnamefont {M.}~\bibnamefont
  {Serbyn}}, \bibinfo {author} {\bibfnamefont {D.~A.}\ \bibnamefont {Abanin}},
  \ and\ \bibinfo {author} {\bibfnamefont {Z.}~\bibnamefont {Papi{\'{c}}}},\
  }\href {\doibase 10.1038/s41567-021-01230-2} {\bibfield  {journal} {\bibinfo
  {journal} {Nature Physics}\ }\textbf {\bibinfo {volume} {17}},\ \bibinfo
  {pages} {675} (\bibinfo {year} {2021})}\BibitemShut {NoStop}%
\bibitem [{\citenamefont {Keating}\ \emph {et~al.}(2006)\citenamefont
  {Keating}, \citenamefont {Mezzadri},\ and\ \citenamefont
  {Novaes}}]{Keating2006}%
  \BibitemOpen
  \bibfield  {author} {\bibinfo {author} {\bibfnamefont {J.~P.}\ \bibnamefont
  {Keating}}, \bibinfo {author} {\bibfnamefont {F.}~\bibnamefont {Mezzadri}}, \
  and\ \bibinfo {author} {\bibfnamefont {M.}~\bibnamefont {Novaes}},\ }\href
  {\doibase 10.1103/physreva.74.012311} {\bibfield  {journal} {\bibinfo
  {journal} {Physical Review A}\ }\textbf {\bibinfo {volume} {74}},\ \bibinfo
  {pages} {012311} (\bibinfo {year} {2006})}\BibitemShut {NoStop}%
\bibitem [{\citenamefont {Chen}\ \emph
  {et~al.}(2006{\natexlab{a}})\citenamefont {Chen}, \citenamefont {Zanardi},
  \citenamefont {Wang},\ and\ \citenamefont {Zhang}}]{Chen2006}%
  \BibitemOpen
  \bibfield  {author} {\bibinfo {author} {\bibfnamefont {Y.}~\bibnamefont
  {Chen}}, \bibinfo {author} {\bibfnamefont {P.}~\bibnamefont {Zanardi}},
  \bibinfo {author} {\bibfnamefont {Z.~D.}\ \bibnamefont {Wang}}, \ and\
  \bibinfo {author} {\bibfnamefont {F.~C.}\ \bibnamefont {Zhang}},\ }\href
  {\doibase 10.1088/1367-2630/8/6/097} {\bibfield  {journal} {\bibinfo
  {journal} {New Journal of Physics}\ }\textbf {\bibinfo {volume} {8}},\
  \bibinfo {pages} {97} (\bibinfo {year} {2006}{\natexlab{a}})}\BibitemShut
  {NoStop}%
\bibitem [{\citenamefont {Chen}\ \emph
  {et~al.}(2006{\natexlab{b}})\citenamefont {Chen}, \citenamefont {Wang},\ and\
  \citenamefont {Zhang}}]{Chen2006a}%
  \BibitemOpen
  \bibfield  {author} {\bibinfo {author} {\bibfnamefont {Y.}~\bibnamefont
  {Chen}}, \bibinfo {author} {\bibfnamefont {Z.~D.}\ \bibnamefont {Wang}}, \
  and\ \bibinfo {author} {\bibfnamefont {F.~C.}\ \bibnamefont {Zhang}},\ }\href
  {\doibase 10.1103/physrevb.73.224414} {\bibfield  {journal} {\bibinfo
  {journal} {Physical Review B}\ }\textbf {\bibinfo {volume} {73}},\ \bibinfo
  {pages} {224414} (\bibinfo {year} {2006}{\natexlab{b}})}\BibitemShut
  {NoStop}%
\bibitem [{\citenamefont {Igl{\'{o}}i}\ and\ \citenamefont
  {Juh{\'{a}}sz}(2008)}]{Igloi2008}%
  \BibitemOpen
  \bibfield  {author} {\bibinfo {author} {\bibfnamefont {F.}~\bibnamefont
  {Igl{\'{o}}i}}\ and\ \bibinfo {author} {\bibfnamefont {R.}~\bibnamefont
  {Juh{\'{a}}sz}},\ }\href {\doibase 10.1209/0295-5075/81/57003} {\bibfield
  {journal} {\bibinfo  {journal} {{EPL} (Europhysics Letters)}\ }\textbf
  {\bibinfo {volume} {81}},\ \bibinfo {pages} {57003} (\bibinfo {year}
  {2008})}\BibitemShut {NoStop}%
\bibitem [{\citenamefont {Igl{\'{o}}i}\ and\ \citenamefont
  {Peschel}(2010)}]{Igloi2010}%
  \BibitemOpen
  \bibfield  {author} {\bibinfo {author} {\bibfnamefont {F.}~\bibnamefont
  {Igl{\'{o}}i}}\ and\ \bibinfo {author} {\bibfnamefont {I.}~\bibnamefont
  {Peschel}},\ }\href {\doibase 10.1209/0295-5075/89/40001} {\bibfield
  {journal} {\bibinfo  {journal} {{EPL} (Europhysics Letters)}\ }\textbf
  {\bibinfo {volume} {89}},\ \bibinfo {pages} {40001} (\bibinfo {year}
  {2010})}\BibitemShut {NoStop}%
\bibitem [{\citenamefont {He}\ \emph {et~al.}(2017)\citenamefont {He},
  \citenamefont {Mag{\'{a}}n},\ and\ \citenamefont {Vandoren}}]{He2017}%
  \BibitemOpen
  \bibfield  {author} {\bibinfo {author} {\bibfnamefont {T.}~\bibnamefont
  {He}}, \bibinfo {author} {\bibfnamefont {J.~M.}\ \bibnamefont {Mag{\'{a}}n}},
  \ and\ \bibinfo {author} {\bibfnamefont {S.}~\bibnamefont {Vandoren}},\
  }\href {\doibase 10.1103/physrevb.95.035130} {\bibfield  {journal} {\bibinfo
  {journal} {Physical Review B}\ }\textbf {\bibinfo {volume} {95}},\ \bibinfo
  {pages} {035130} (\bibinfo {year} {2017})}\BibitemShut {NoStop}%
\bibitem [{\citenamefont {Rolandi}\ and\ \citenamefont
  {Wilming}(2020)}]{Rolandi2020}%
  \BibitemOpen
  \bibfield  {author} {\bibinfo {author} {\bibfnamefont {A.}~\bibnamefont
  {Rolandi}}\ and\ \bibinfo {author} {\bibfnamefont {H.}~\bibnamefont
  {Wilming}},\ }\href@noop {} {\  (\bibinfo {year} {2020})},\ \Eprint
  {http://arxiv.org/abs/2008.11764} {arXiv:2008.11764 [quant-ph]} \BibitemShut
  {NoStop}%
\bibitem [{\citenamefont {Haferkamp}\ \emph {et~al.}(2021)\citenamefont
  {Haferkamp}, \citenamefont {Bertoni}, \citenamefont {Roth},\ and\
  \citenamefont {Eisert}}]{Haferkamp2021}%
  \BibitemOpen
  \bibfield  {author} {\bibinfo {author} {\bibfnamefont {J.}~\bibnamefont
  {Haferkamp}}, \bibinfo {author} {\bibfnamefont {C.}~\bibnamefont {Bertoni}},
  \bibinfo {author} {\bibfnamefont {I.}~\bibnamefont {Roth}}, \ and\ \bibinfo
  {author} {\bibfnamefont {J.}~\bibnamefont {Eisert}},\ }\href {\doibase
  10.1103/prxquantum.2.040308} {\bibfield  {journal} {\bibinfo  {journal}
  {{PRX} Quantum}\ }\textbf {\bibinfo {volume} {2}},\ \bibinfo {pages} {040308}
  (\bibinfo {year} {2021})}\BibitemShut {NoStop}%
\bibitem [{\citenamefont {Wilming}\ \emph {et~al.}(2019)\citenamefont
  {Wilming}, \citenamefont {Goihl}, \citenamefont {Roth},\ and\ \citenamefont
  {Eisert}}]{Wilming2019}%
  \BibitemOpen
  \bibfield  {author} {\bibinfo {author} {\bibfnamefont {H.}~\bibnamefont
  {Wilming}}, \bibinfo {author} {\bibfnamefont {M.}~\bibnamefont {Goihl}},
  \bibinfo {author} {\bibfnamefont {I.}~\bibnamefont {Roth}}, \ and\ \bibinfo
  {author} {\bibfnamefont {J.}~\bibnamefont {Eisert}},\ }\href {\doibase
  10.1103/physrevlett.123.200604} {\bibfield  {journal} {\bibinfo  {journal}
  {Physical Review Letters}\ }\textbf {\bibinfo {volume} {123}},\ \bibinfo
  {pages} {200604} (\bibinfo {year} {2019})}\BibitemShut {NoStop}%
\bibitem [{\citenamefont {Botero}\ and\ \citenamefont
  {Reznik}(2004)}]{Botero2004}%
  \BibitemOpen
  \bibfield  {author} {\bibinfo {author} {\bibfnamefont {A.}~\bibnamefont
  {Botero}}\ and\ \bibinfo {author} {\bibfnamefont {B.}~\bibnamefont
  {Reznik}},\ }\href {\doibase 10.1016/j.physleta.2004.08.037} {\bibfield
  {journal} {\bibinfo  {journal} {Physics Letters A}\ }\textbf {\bibinfo
  {volume} {331}},\ \bibinfo {pages} {39} (\bibinfo {year} {2004})}\BibitemShut
  {NoStop}%
\bibitem [{\citenamefont {Shapourian}\ \emph {et~al.}(2017)\citenamefont
  {Shapourian}, \citenamefont {Shiozaki},\ and\ \citenamefont
  {Ryu}}]{Shapourian2017}%
  \BibitemOpen
  \bibfield  {author} {\bibinfo {author} {\bibfnamefont {H.}~\bibnamefont
  {Shapourian}}, \bibinfo {author} {\bibfnamefont {K.}~\bibnamefont
  {Shiozaki}}, \ and\ \bibinfo {author} {\bibfnamefont {S.}~\bibnamefont
  {Ryu}},\ }\href {\doibase 10.1103/physrevb.95.165101} {\bibfield  {journal}
  {\bibinfo  {journal} {Physical Review B}\ }\textbf {\bibinfo {volume} {95}},\
  \bibinfo {pages} {165101} (\bibinfo {year} {2017})}\BibitemShut {NoStop}%
\bibitem [{\citenamefont {Eisert}\ \emph {et~al.}(2018)\citenamefont {Eisert},
  \citenamefont {Eisler},\ and\ \citenamefont {Zimbor{\'{a}}s}}]{Eisert2018}%
  \BibitemOpen
  \bibfield  {author} {\bibinfo {author} {\bibfnamefont {J.}~\bibnamefont
  {Eisert}}, \bibinfo {author} {\bibfnamefont {V.}~\bibnamefont {Eisler}}, \
  and\ \bibinfo {author} {\bibfnamefont {Z.}~\bibnamefont {Zimbor{\'{a}}s}},\
  }\href {\doibase 10.1103/physrevb.97.165123} {\bibfield  {journal} {\bibinfo
  {journal} {Physical Review B}\ }\textbf {\bibinfo {volume} {97}},\ \bibinfo
  {pages} {165123} (\bibinfo {year} {2018})}\BibitemShut {NoStop}%
\bibitem [{\citenamefont {Vidal}\ \emph {et~al.}(2003)\citenamefont {Vidal},
  \citenamefont {Latorre}, \citenamefont {Rico},\ and\ \citenamefont
  {Kitaev}}]{Vidal2003}%
  \BibitemOpen
  \bibfield  {author} {\bibinfo {author} {\bibfnamefont {G.}~\bibnamefont
  {Vidal}}, \bibinfo {author} {\bibfnamefont {J.~I.}\ \bibnamefont {Latorre}},
  \bibinfo {author} {\bibfnamefont {E.}~\bibnamefont {Rico}}, \ and\ \bibinfo
  {author} {\bibfnamefont {A.}~\bibnamefont {Kitaev}},\ }\href {\doibase
  10.1103/physrevlett.90.227902} {\bibfield  {journal} {\bibinfo  {journal}
  {Physical Review Letters}\ }\textbf {\bibinfo {volume} {90}},\ \bibinfo
  {pages} {227902} (\bibinfo {year} {2003})}\BibitemShut {NoStop}%
\bibitem [{\citenamefont {Chapman}\ and\ \citenamefont
  {Flammia}(2020)}]{Chapman2020}%
  \BibitemOpen
  \bibfield  {author} {\bibinfo {author} {\bibfnamefont {A.}~\bibnamefont
  {Chapman}}\ and\ \bibinfo {author} {\bibfnamefont {S.~T.}\ \bibnamefont
  {Flammia}},\ }\href {\doibase 10.22331/q-2020-06-04-278} {\bibfield
  {journal} {\bibinfo  {journal} {Quantum}\ }\textbf {\bibinfo {volume} {4}},\
  \bibinfo {pages} {278} (\bibinfo {year} {2020})}\BibitemShut {NoStop}%
\bibitem [{\citenamefont {Elman}\ \emph {et~al.}(2021)\citenamefont {Elman},
  \citenamefont {Chapman},\ and\ \citenamefont {Flammia}}]{Elman2021}%
  \BibitemOpen
  \bibfield  {author} {\bibinfo {author} {\bibfnamefont {S.~J.}\ \bibnamefont
  {Elman}}, \bibinfo {author} {\bibfnamefont {A.}~\bibnamefont {Chapman}}, \
  and\ \bibinfo {author} {\bibfnamefont {S.~T.}\ \bibnamefont {Flammia}},\
  }\href {\doibase 10.1007/s00220-021-04220-w} {\bibfield  {journal} {\bibinfo
  {journal} {Communications in Mathematical Physics}\ }\textbf {\bibinfo
  {volume} {388}},\ \bibinfo {pages} {969} (\bibinfo {year}
  {2021})}\BibitemShut {NoStop}%
\bibitem [{\citenamefont {Alba}\ \emph {et~al.}(2010)\citenamefont {Alba},
  \citenamefont {Tagliacozzo},\ and\ \citenamefont {Calabrese}}]{Alba2010}%
  \BibitemOpen
  \bibfield  {author} {\bibinfo {author} {\bibfnamefont {V.}~\bibnamefont
  {Alba}}, \bibinfo {author} {\bibfnamefont {L.}~\bibnamefont {Tagliacozzo}}, \
  and\ \bibinfo {author} {\bibfnamefont {P.}~\bibnamefont {Calabrese}},\ }\href
  {\doibase 10.1103/physrevb.81.060411} {\bibfield  {journal} {\bibinfo
  {journal} {Physical Review B}\ }\textbf {\bibinfo {volume} {81}},\ \bibinfo
  {pages} {060411} (\bibinfo {year} {2010})}\BibitemShut {NoStop}%
\bibitem [{\citenamefont {Zirnbauer}(2021)}]{Zirnbauer2021}%
  \BibitemOpen
  \bibfield  {author} {\bibinfo {author} {\bibfnamefont {M.~R.}\ \bibnamefont
  {Zirnbauer}},\ }\href {\doibase 10.1063/5.0035358} {\bibfield  {journal}
  {\bibinfo  {journal} {Journal of Mathematical Physics}\ }\textbf {\bibinfo
  {volume} {62}},\ \bibinfo {pages} {021101} (\bibinfo {year}
  {2021})}\BibitemShut {NoStop}%
\bibitem [{Note1()}]{Note1}%
  \BibitemOpen
  \bibinfo {note} {We are aware that strictly speaking there is no
  tensor-product structure. But there is no harm in writing it like
  this.}\BibitemShut {Stop}%
\bibitem [{\citenamefont {Altland}\ and\ \citenamefont
  {Zirnbauer}(1997)}]{Altland1997}%
  \BibitemOpen
  \bibfield  {author} {\bibinfo {author} {\bibfnamefont {A.}~\bibnamefont
  {Altland}}\ and\ \bibinfo {author} {\bibfnamefont {M.~R.}\ \bibnamefont
  {Zirnbauer}},\ }\href {\doibase 10.1103/physrevb.55.1142} {\bibfield
  {journal} {\bibinfo  {journal} {Physical Review B}\ }\textbf {\bibinfo
  {volume} {55}},\ \bibinfo {pages} {1142} (\bibinfo {year}
  {1997})}\BibitemShut {NoStop}%
\bibitem [{\citenamefont {Heinzner}\ \emph {et~al.}(2005)\citenamefont
  {Heinzner}, \citenamefont {Huckleberry},\ and\ \citenamefont
  {Zirnbauer}}]{Heinzner2005}%
  \BibitemOpen
  \bibfield  {author} {\bibinfo {author} {\bibfnamefont {P.}~\bibnamefont
  {Heinzner}}, \bibinfo {author} {\bibfnamefont {A.}~\bibnamefont
  {Huckleberry}}, \ and\ \bibinfo {author} {\bibfnamefont {M.}~\bibnamefont
  {Zirnbauer}},\ }\href {\doibase 10.1007/s00220-005-1330-9} {\bibfield
  {journal} {\bibinfo  {journal} {Communications in Mathematical Physics}\
  }\textbf {\bibinfo {volume} {257}},\ \bibinfo {pages} {725} (\bibinfo {year}
  {2005})}\BibitemShut {NoStop}%
\bibitem [{\citenamefont {Kitaev}\ \emph {et~al.}(2009)\citenamefont {Kitaev},
  \citenamefont {Lebedev},\ and\ \citenamefont {Feigel'man}}]{Kitaev2009}%
  \BibitemOpen
  \bibfield  {author} {\bibinfo {author} {\bibfnamefont {A.}~\bibnamefont
  {Kitaev}}, \bibinfo {author} {\bibfnamefont {V.}~\bibnamefont {Lebedev}}, \
  and\ \bibinfo {author} {\bibfnamefont {M.}~\bibnamefont {Feigel'man}},\ }in\
  \href {\doibase 10.1063/1.3149495} {\emph {\bibinfo {booktitle} {{AIP}
  Conference Proceedings}}}\ (\bibinfo  {publisher} {{AIP}},\ \bibinfo {year}
  {2009})\BibitemShut {NoStop}%
\bibitem [{\citenamefont {Ryu}\ \emph {et~al.}(2010)\citenamefont {Ryu},
  \citenamefont {Schnyder}, \citenamefont {Furusaki},\ and\ \citenamefont
  {Ludwig}}]{Ryu2010}%
  \BibitemOpen
  \bibfield  {author} {\bibinfo {author} {\bibfnamefont {S.}~\bibnamefont
  {Ryu}}, \bibinfo {author} {\bibfnamefont {A.~P.}\ \bibnamefont {Schnyder}},
  \bibinfo {author} {\bibfnamefont {A.}~\bibnamefont {Furusaki}}, \ and\
  \bibinfo {author} {\bibfnamefont {A.~W.~W.}\ \bibnamefont {Ludwig}},\ }\href
  {\doibase 10.1088/1367-2630/12/6/065010} {\bibfield  {journal} {\bibinfo
  {journal} {New Journal of Physics}\ }\textbf {\bibinfo {volume} {12}},\
  \bibinfo {pages} {065010} (\bibinfo {year} {2010})}\BibitemShut {NoStop}%
\bibitem [{\citenamefont {Ludwig}(2015)}]{Ludwig2015}%
  \BibitemOpen
  \bibfield  {author} {\bibinfo {author} {\bibfnamefont {A.~W.~W.}\
  \bibnamefont {Ludwig}},\ }\href {\doibase 10.1088/0031-8949/2015/t168/014001}
  {\bibfield  {journal} {\bibinfo  {journal} {Physica Scripta}\ }\textbf
  {\bibinfo {volume} {T168}},\ \bibinfo {pages} {014001} (\bibinfo {year}
  {2015})}\BibitemShut {NoStop}%
\end{thebibliography}
%

\appendix
\section{Proof of Lemma~\ref{lemma:basic}}
The core of the proof Lemma~\ref{lemma:basic} is essentially given by the singular value decomposition, but we provide a somewhat pedestrian argument. In the following, let us write $\tilde h = h_{AB}$. 
Any normalized eigenstate of $h$ may be written as $\psi_k = \frac{1}{\sqrt 2}(\psi_{A,k},\psi_{B,k})^\top$.
Given~\eqref{eq:sublattice}, the eigenvalue equation for $\psi_k$ reads
\begin{align}\label{eq:eigenvalues}
	\tilde h \psi_{B,k} = \eps_k \psi_{A,k},\quad \tilde h^\dagger \psi_{A,k} = \eps_k \psi_{B,k}. 
\end{align}
We first consider the case $\eps_k\neq 0$. 
Eq.~\eqref{eq:eigenvalues} then implies that $\psi_{B,k}$ has to be an eigenvector of $\tilde h^\dagger \tilde h$
and $\psi_{A,k}$ has to be an eigenvector of $\tilde h\tilde h^\dagger$, both with eigenvalue $\eps_k^2$. Using again~\eqref{eq:eigenvalues} we find that $\psi_{A,k}$ and $\psi_{B,k}$ have the same norm (and are hence related by an isometry). 
Since $\psi_k$ is normalized, so are the $\psi_{A,k}$ and $\psi_{B,k}$ and because the $\psi_k$ have to be orthogonal for different values of $k$, the same is true for the set of $\psi_{A,k}$ and the set of $\psi_{B,k}$.
We thus obtain an orthonormal basis for the subspace of $\mathbb C^{|B|}$ orthogonal to the nullspace of $\tilde h$ and an isometric subspace in $\mathbb C^{|A|}$ spanned by the $\psi_{A,k}$. 
We can now first complete the $\psi_{B,k}$ to an orthonormal basis on $\mathbb C^{|B|}$ using vectors in the nullspace of $\tilde h$ with associated $\psi_{A,k}$ that also have to be orthogonal to the previous $\psi_{A,k}$. The $\psi_{A,k}$ then provide an isometric embedding of $\mathbb C^{|B|}$ into $\mathbb C^{|A|}$ and we may write them as $V\psi_{B,k}$ for some isometry $V$.
We have so far obtained the first $k=1,\ldots,|B|$ solutions to \eqref{eq:eigenvalues}. By flipping the signs of the $B$-components, we obtain $|B|$ more solutions, which we label by $k=|B|+1,\ldots,2|B|$.
There remain $|A|-|B|$ solutions $\psi_k$ with $k=2|B|+l$ and $l=1,\ldots,|A|-|B|$. Since they have to be orthogonal to the already constructed solutions we find for $k\leq |B|$
\begin{align}
	0=\bra{\psi_{2|B|+l}}\left(\ket{\psi_k} + \ket{\psi_{k+|B|}}\right) = \sqrt{2}\braket{\psi_{B,2|B|+l}}{\psi_{B,k}}.\nonumber
\end{align}
Since the $\psi_{B,k}$ for $k=1,\ldots, B$ form a basis, we find that $\psi_{B,2|B|+l}=0$. 
Hence the remaining solutions are supported on $A$ only and can be written as $\psi_{2|B|+l}=(\phi_l,0)^\top$. 
This finishes the proof.

\section{Non-number-conserving case}
We now study a general Hamiltonian of non-interacting fermions, which may also include \emph{pairing terms} of the form $f_i f_j + f_j^\dagger f_i^\dagger$. 
Such Hamiltonians do not preserve the total number of Fermions $\hat N = \sum_j f_j^\dagger f_j$. 
We will now first introduce the necessary background material to conveniently work with such Hamiltonians \cite{Vidal2003}.
This is best done by introducing \emph{Majorana operators} as
\begin{align}
	\omega_{2j-1} = f^\dagger_j + f_j,\quad \omega_{2j} = -\mathrm{i}(f_j^\dagger - f_j).
\end{align}
Thus the Majorana operators are twice the Hermitian and Anti-Hermitian part of $f_j^\dagger$, respectively.
They fulfill the anti-commutation relations
\begin{align}
	\{\omega_j, \omega_k\} = 2\delta_{j,k}.
\end{align}
Any Hamiltonian that is quadratic in the $f_j,f_k^\dagger$ can be expressed in the form
\begin{align}
	H_M = \frac{\mathrm i}{4} \omega^\top h_M \omega,
\end{align}
where now $h_M=-h_M^\top$ is a \emph{real}, anti-symmetric $2N\times 2N$ matrix and $\omega = (\omega_1,\omega_2,\ldots,\omega_{2N})^\top$ is a vector collecting the Majorana operators.  
As a real and anti-symmetric  $2N\times 2N$-matrix, $h$ may be written as 
\begin{align}\label{eq:normalform}
	h_M = R^\top \left[\bigoplus_{k=1}^N \eps_k \begin{pmatrix} 0&-1\\ 1&0\end{pmatrix}\right] R,\quad \eps_k \geq 0,
\end{align}
where $R$ is a real orthogonal matrix. We will see below how to construct $R$ explicitly. 
If we define the new Majorana operators
\begin{align}
		\omega'_k := \sum_{k} R_{kj}\omega_j
\end{align}
and corresponding fermionic annihilation and creation operators 
\begin{align}
	a^\dagger_k &= \frac{1}{2}\left(\omega'_{2k-1} + \mathrm{i} \omega'_{2k}\right),\\ 
		a_k &= \frac{1}{2}\left(\omega'_{2k-1} - \mathrm{i} \omega'_{2k}\right),
\end{align}
then the Hamiltonian takes the form
\begin{align}
	H_M = \sum_k \eps_k \left(a^\dagger_k a_k -\frac{1}{2}\right).
\end{align}
Since all $\eps_k\geq 0$ (by convention), the ground-state $\ket{E_{\mathrm{GS}}}$ fulfills $a_k \ket{E_{\mathrm{GS}}}=0$ for all $k$. 
This vacuum for the $a_k$ is in general \emph{not} given by the vacuum $\ket{0}$ of the $f_j$. 
In case some of the $\eps_k=0$, additional ground states may be obtained by acting with the $a_k^\dagger$ on $\ket{E_{\mathrm{GS}}}$.

Let $r_j$ denote the $j$-th column of $R^\top$. Then \eqref{eq:normalform} tells us that
\begin{align}\label{eq:symplectic_evs}
		h_M r_{2k-1} = \eps_k r_{2k},\quad h_M r_{2k} = -\eps_k r_{2k-1}.
\end{align}
Consequently, the \emph{complex} vectors defined as 
\begin{align}
	\psi_{k}^\pm := \frac{1}{\sqrt 2}\left(r_{2k-1} \mp \mathrm{i} r_{2k}\right),  
\end{align}
with $\psi_k^+ = \overline{\psi_k^-}$ fulfill
\begin{align}
		-\mathrm{i} h \psi_k^\pm = \pm \eps_k\psi_k^\pm.
\end{align}
Conversely, if $\psi_k$ fulfills $-\mathrm{i} h_M \psi_k = \eps_k$ with $\eps_k\geq 0$, then we can define corresponding vectors fulfilling \eqref{eq:symplectic_evs} by taking their real and imaginary parts, respectively:
\begin{align}
		r_{2k-1} = \frac{1}{\sqrt{2}}\left(\psi_k + \overline{\psi_k}\right),\quad r_{2k} = \frac{\mathrm i}{\sqrt{2}}\left(\psi_k - \overline{\psi_k}\right).
\end{align}
Note that if $\psi_k$ is an eigenvector of $-\mathrm i h_M$ with eigenvalue $\eps_k$,  then $\overline{\psi_k}$ is an eigenvector with eigenvalue $-\eps_k$. In particular, $\psi_k$ and $\overline{\psi_k}$ are orthogonal with respect to the complex inner product,
\begin{align}\label{eq:inner_products}
	\psi_k^\dagger\psi_j = \delta_{jk} ,\quad  \overline{\psi_k}^\dagger \psi_k =\psi_k^\top \psi_k= 0.
\end{align}

The normal form \eqref{eq:normalform} may now be obtained by first obtaining the $N$ eigenvectors of $-\mathrm i h_M$ with positive (semi-definite) eigenvalues and then constructing the $r_{2k-1},r_{2k}$ from them. 
In terms of the $r$-vectors, \eqref{eq:normalform} can be expressed as
\begin{align}
	h_M = \sum_{k=1}^N \eps_k \left[r_{2k} r_{2k-1}^\top - r_{2k-1} r_{2k}^\top\right].
\end{align}
An important property of quadratic Hamiltonians is that their eigenstates are so-called \emph{Gaussian states}, that are completely determined by their real, anti-symmetric \emph{covariance matrix} defined as
\begin{align}
	\Gamma_{jk} = \frac{\mathrm i}{2}\Tr\left(\rho [\omega_j,\omega_k]\right).
\end{align}
As real, anti-symmetric matrix $\Gamma$ has a similar structure as $h$, with values $\nu_k\leq 1$ replacing the $\eps_k$. 
We will call the $\nu_k$ the \emph{mode-spectrum} of $\Gamma$. 
A Gaussian state is pure if and only if $\nu_k=1$ for all $k$.
Importantly, the covariance matrix of the reduced state $\rho_B$ is simply given by the restriction $\Gamma_B$ of $\Gamma$ with $j,k$ corresponding to the subsystem $B$.
If $\Gamma_B$ has mode-spectrum $\nu_k$ with $k=1,\ldots,|B|$, then its von Neumann entropy is given by
\begin{align}
	S(B)_\rho = \sum_k H_2\left(\frac{1}{2}(1+\nu_k)\right), 
\end{align}
with $H_2(x) = -x \log_2(x) - (1-x)\log_2(1-x)$ the binary entropy.
In particular, a pure Gaussian state is maximally entangled between $A$ and its complement $B$ if and only if $\nu_k=0$ for all $k=1,\ldots,|B|$, which happens if and only if $\Gamma_B=0$ (we remind the reader that we assume $|A|\geq |B|$. Then $\Gamma_B=0$ does not imply $\Gamma_A=0$).

Given the normal form \eqref{eq:normalform}, the pure energy-eigenstates $\ket{E_s}$ have covariance matrices
\begin{align}\label{eq:coveigen}
	\Gamma_s = -\sum_{k=1}^N (-1)^{s_k} \left[r_{2k} r_{2k-1}^\top - r_{2k-1} r_{2k}^\top\right],
\end{align}
where $s_k \in \{0,1\}$, with energies
\begin{align}
	E_s = -\frac{1}{2}\sum_{k=1}^N (-1)^{s_k} \eps_k.	
\end{align}
We now define the sublattice symmetry on the level of Majorana operators.
\begin{definition}[Generalized sublattice symmetry] We will say that $H_M$ has sublattice symmetry if there exists two even and disjoint sets $A,B$ such that $A\cup B=\{1,\ldots,2N\}$ and $h_M$ couples the \emph{Majorana operators} $\omega_j$ with $j\in A$ only with those of $B$ and vice-versa. 
\end{definition}
The physical meaning of the sublattice symmetry only becomes clear once we construct again fermionic modes from the Majorana operators in the two sets $A$ and $B$.
The statement that we will now proof is that the fermionic modes constructed from the Majorana operators in $B$ are maximally entangled with those constructed from $A$.
The sublattice symmetry implies that the Hermitian matrix $-\mathrm{i} h_M$ is of the form
\begin{align}
	-\mathrm i h_M = 
	\begin{pmatrix} 
		0& -\mathrm i h_{M,AB} \\
		\mathrm{i} h_{M,AB}^\top & 0
	\end{pmatrix}
	=
	\begin{pmatrix} 
		0& \tilde h_M \\
		\tilde h^\dagger_M & 0
	\end{pmatrix},
\end{align}
where $h_{AB}$ and $\tilde{h}_M=-\mathrm i h_{M,AB}$ are now $|A|\times |B|$-matrices.
As before we need to determine the eigenvectors $\psi_k = \frac{1}{\sqrt{2}}(\psi_{A,k},\psi_{B,k})^\top$ of $-\mathrm i h_M$ with eigenvalues $\eps_k\geq 0$. Writing out the eigenvalue equation in terms of the $A$ and $B$ components, we get again
\begin{align}
	\tilde h_M \psi_{B,k} = \eps_k \psi_{A,k},\quad \tilde h_M^\dagger \psi_{A,k} = \eps_k \psi_{B,k}.
\end{align}
As a real matrix, $h_{M,AB}$ can be written using the real singular value decomposition as 
\begin{align}
h_{M,AB} = V' D W^\top,
\end{align}
where $W$ is an orthogonal $|B|\times |B|$ matrix, $D$ is a real, positive semi-definite and diagonal $|B|\times |B|$-matrix and $V$ is a real isometric $|A|\times |B|$-matrix.
For now denote the diagonal entries of $D$ as $D_{kk} =: d_k$. Let us define $V:= -\mathrm{i} V'W^\top$ and set
\begin{align}
	\psi_{B,k} = W \vec e_k,\quad \psi_{A,k} = V\psi_{B,k} = -\mathrm i V' \vec e_k,
\end{align}
with $\vec e_k$ the $k$-th unit vector on $\CC^{|B|}$. We can easily see that these provide a solution to the eigenvalue equation with $\eps_k=d_k$:
\begin{align}
	\tilde{h}_M \psi_{B,k} &= -\mathrm{i} V'D W^\top W \vec e_k = d_k(-\mathrm i  V'e_k) = d_k \psi_{A,k}
\end{align}
and similarly for the second equation. Thus we have obtain $|B|$ solutions with $\eps_k\geq 0$.
The vectors 
\begin{align}
	\psi_{k+|B|}:= \overline{\psi_k}
\end{align}
are eigenvectors of $-\mathrm i h_M$ with eigenvalues $\eps_{k+|B|}=-\eps_k$. 
The remaining $|A|-|B|$ many eigenvectors of $-\mathrm{i} h_M$ take the form
\begin{align}
	\psi_{2|B|+l} = (\phi_{A,l},0)^\top,\quad l=1,\ldots,|A|-|B|,
\end{align}
where the $\phi_{A,l}$ are an orthonormal basis of the subspace of $\CC^{|A|}$ orthogonal to $V\mathbb C^{|B|}$.
They are part of the nullspace of $-\mathrm i h_M$. 

Having found the $N$ eigenvectors of $-\mathrm i h_M$ with non-negative eigenvalues, we can now construct the corresponding vectors $r_{2k-1},r_{2k}$.
We find 
\begin{align}
	(r_{2k}r_{2k-1}^\top - r_{2k-1}r_{2k}^\top)|_B 
	&=\frac{\mathrm{i}}{2}\left[\psi_{B,k} \psi_{B,k}^\dagger - \overline{\psi_{B,k} \psi_{B,k}^\dagger} \right]\nonumber\\
	&=\frac{\mathrm{i}}{2}\left[P_k - \overline{P_k}\right],
\end{align}
where we defined the orthogonal rank-1 projectors $P_k = \psi_{B,k} \psi_{B,k}^\dagger$.
Note that for $k=2|B|+l$ we have $P_k=0$ since $\psi_{B,k}=0$. 
Thus, the covariance matrix of $\ket{E_s}$ restricted to $B$ takes the form
\begin{align}
	\Gamma_{s,B} = -\frac{\mathrm i}{2}\sum_{k=1}^{|B|} (-1)^{s_k} \left[P_k - \overline{P_k} \right].
\end{align}
Since we chose $\psi_{B,k}$ to be real vectors, we have $P_k = \overline{P_k}$ and hence $\Gamma_{s,B}=0$, which is equivalent to maximum entanglement as discussed above. QED.

We close by discussing in more detail how this result relates to  the number-conserving case.
In our notational convention, the fermionic number operator can be written, up to a constant shift, as 
\begin{align}
	\hat N = \sum_{i=1}^N f_i^\dagger f_i = \frac{\mathrm i}{2} \omega^\top n \omega + {(\mathrm{const.})}\id,
\end{align}
where
\begin{align}
n=\id_N \otimes J_2 := \id_N \otimes \begin{pmatrix} 0 & -1\\ 1&0\end{pmatrix}.
\end{align}
The number operator commutes with the Hamiltonian $H_F$ if and only if $[h_M,n]=0$. 
Since $h_M$ is a real, anti-symmetric matrix, this implies that it is of the form
\begin{align}
h_M = a\otimes \id_2 + b \otimes J_2,
\end{align}
with $a=-a^\top$ and $b=b^\top$ real $N\times N$ matrices.
If we additionally impose sublattice symmetry, this implies that all the $\eps_k\geq 0$ are at least doubly degenerate, which leaves us (at least) with the freedom to choose a basis in the (at least) 2-dimensional eigenspace of $-\mathrm i h_M$ with eigenvalue $\eps_k$. 

To obtain eigenstates that are also eigenstates of the number operator, the $\psi_k$ have to be eigenstates of $n$. 
This in turn implies that $\psi_{B,k}$ has to to be an eigenstate of $n_B = \id_{|B|/2}\otimes J_2$, 
which in turn implies that $\psi_{B,k}$ cannot be real up to a phase, since $n_B$ is an anti-symmetric matrix. 
Thus, if we want to obtain energy eigenstates that are also eigenstates of the number operator, we cannot choose $\psi_{B,k}$ to be real up to a phase, which implies that $P_k \neq \overline{P_k}$.
However, it turns out that we can choose $\psi_{B,k}$ so that $\psi_{B,k+|B|/2} = \overline{\psi_{B,k}}$ with $k=1,\ldots,|B|/2$. 
I.e., the $\psi_{B,k}$ come in pairs of complex conjugates.
Therefore we find
\begin{align}
	\Gamma_{s,B} = -\frac{\mathrm i}{2}\sum_{k=1}^{|B|/2} \left((-1)^{s_k}-(-1)^{s_{k+|B|/2}}\right) \left[P_k - \overline{P_k} \right].
\end{align}
We can now analyse the mode-spectrum of $\Gamma_{s,B}$, which is simply (up to a doubling) given by the singular values of $\Gamma_{s,B}$.
Since the terms of the sum are supported on different, orthogonal $2$-dimensional subspaces, we can consider each term separately.
If $s_k=s_{k+|B|/2}$ (i.e., $s_k\oplus s_{k+|B|/2}=0$), the corresponding term vanishes, yielding $\nu_k=0$.
This corresponds to one singlet of entanglement between $A$ and $B$.
Conversely, if $s_k\oplus s_{k+|B|/2}=0$, the corresponding term has eigenvalues $\pm\mathrm{i}$, yielding $\nu_k=1$.
This corresponds to a product state between $A$ and $B$.
Thus, the total entanglement is given by
\begin{align}
	\tilde n(s) = \sum_{k=1}^{|B|/2} s_k\oplus s_{k+|B|/2}\oplus 1
\end{align}
singlets. (Remember that $B$ contains an even number of Majorana operators that correspond to $|B|/2$ many fermions. Hence maximum entanglement is given by $|B|/2$ singlets.)

We close by expressing the freedom in defining the $\psi_{k}$ in the number-conserving case in yet different words: To obtain a normal-form of $h$ as in \eqref{eq:normalform} we are allowed to choose $R$ from the orthogonal group $O(2N)$. However, if we do not impose number-conservation, then the normal-form is certainly not unique. If we want to preserve number-conservation, then the orthogonal transformation $R$ has to be chosen from the \emph{unitary} subgroup $U(N)\subseteq O(2N)$ of matrices $R$ that fulfill $[R,n]=0$.
The unitary $U$ that diagonalizes the \emph{hermitian} single-particle Hamiltonian $h$ in the number-conserving case (whose rows are the normal modes $\psi_k(i)$) is related to this $R$ as
\begin{align}
	R = \mathrm{Re} U\otimes \id_2 +  \mathrm{Im} U\otimes J_2,
\end{align}
while the Hamiltonian in the Majorana representation takes the form
\begin{align}
	h_M = \mathrm{Im} h \otimes \id_2 + \mathrm{Re} h \otimes J_2.
\end{align}
It can then be readily checked that $R^\top h_M R$ is in normal-form.

\clearpage

\end{document}